\documentclass[%
 twocolumn,
 superscriptaddress,
 amsmath,amssymb,
 aps,
 pra,
]{revtex4-2}

\usepackage{graphicx}
\usepackage{dcolumn}
\usepackage{bm}
\usepackage{color,soul}
\usepackage{url}
\usepackage{xr}

\begin{document}


\title{Direct observation of the exciton polaron by serial femtosecond crystallography on single CsPbBr$_3$ quantum dots }

\author{Zhou Shen}
\affiliation{Max Planck Institute for the Structure and Dynamics of Matter, 22761 Hamburg, Germany}

\author{Margarita Samoli}
\affiliation{Physics and Chemistry of Nanostructures, Ghent University, Gent 9000, Belgium}

\author{Onur Erdem}
\affiliation{Physics and Chemistry of Nanostructures, Ghent University, Gent 9000, Belgium}

\author{Johan Bielecki}
\affiliation{European XFEL, 22869 Schenefeld, Germany}

\author{Amit Kumar Samanta}
\affiliation{Center for Free-Electron Laser Science CFEL, Deutsches Elektronen-Synchrotron DESY, 22607 Hamburg, Germany}

\author{Juncheng E}
\affiliation{European XFEL, 22869 Schenefeld, Germany}

\author{Armando Estillore}
\affiliation{Center for Free-Electron Laser Science CFEL, Deutsches Elektronen-Synchrotron DESY, 22607 Hamburg, Germany}

\author{Chan Kim}
\affiliation{European XFEL, 22869 Schenefeld, Germany}

\author{Yoonhee Kim}
\affiliation{European XFEL, 22869 Schenefeld, Germany}

\author{Jayanath Koliyadu}
\affiliation{European XFEL, 22869 Schenefeld, Germany}

\author{Romain Letrun}
\affiliation{European XFEL, 22869 Schenefeld, Germany}

\author{Federico Locardi}
\affiliation{Dipartimento di Chimica e Chimica Industriale, Università degli Studi di Genova, 16146 Genova, Italy}
\affiliation{Physics and Chemistry of Nanostructures, Ghent University, Gent 9000, Belgium}

\author{Jannik Lübke}
\affiliation{Center for Free-Electron Laser Science CFEL, Deutsches Elektronen-Synchrotron DESY, 22607 Hamburg, Germany}

\author{Abhishek Mall}
\affiliation{Max Planck Institute for the Structure and Dynamics of Matter, 22761 Hamburg, Germany}

\author{Diogo Melo}
\affiliation{European XFEL, 22869 Schenefeld, Germany}

\author{Grant Mills}
\affiliation{European XFEL, 22869 Schenefeld, Germany}

\author{Safi Rafie-Zinedine}
\affiliation{European XFEL, 22869 Schenefeld, Germany}

\author{Adam Round}
\affiliation{European XFEL, 22869 Schenefeld, Germany}

\author{Tokushi Sato}
\affiliation{European XFEL, 22869 Schenefeld, Germany}

\author{Raphael de Wijn}
\affiliation{European XFEL, 22869 Schenefeld, Germany}

\author{Tamme Wollweber}
\affiliation{Max Planck Institute for the Structure and Dynamics of Matter, 22761 Hamburg, Germany}
\affiliation{The Hamburg Center for Ultrafast Imaging, 22761 Hamburg, Germany}

\author{Lena Worbs}
\affiliation{Center for Free-Electron Laser Science CFEL, Deutsches Elektronen-Synchrotron DESY, 22607 Hamburg, Germany}

\author{Yulong Zhuang}
\affiliation{Max Planck Institute for the Structure and Dynamics of Matter, 22761 Hamburg, Germany}

\author{Adrian P. Mancuso}
\affiliation{European XFEL, 22869 Schenefeld, Germany}
\affiliation{Department of Chemistry and Physics, La Trobe Institute for Molecular Science, La Trobe University, Melbourne, VIC 3086, Australia}

\author{Richard Bean}
\affiliation{European XFEL, 22869 Schenefeld, Germany}

\author{Henry N. Chapman}
\affiliation{Center for Free-Electron Laser Science CFEL, Deutsches Elektronen-Synchrotron DESY, 22607 Hamburg, Germany}
\affiliation{The Hamburg Center for Ultrafast Imaging, 22761 Hamburg, Germany}
\affiliation{Department of Physics, University of Hamburg, 22761 Hamburg, Germany}

\author{Jochen Küpper}
\affiliation{Center for Free-Electron Laser Science CFEL, Deutsches Elektronen-Synchrotron DESY, 22607 Hamburg, Germany}
\affiliation{The Hamburg Center for Ultrafast Imaging, 22761 Hamburg, Germany}
\affiliation{Department of Physics, University of Hamburg, 22761 Hamburg, Germany}

\author{Ivan Infante}
\affiliation{BCMaterials, Basque Center for Materials, Applications, and Nanostructures, UPV/EHU Science Park, Leioa, 48940 Spain}
\affiliation{Ikerbasque Basque Foundation for Science, Bilbao 48009, Spain}

\author{Holger Lange}
\affiliation{The Hamburg Center for Ultrafast Imaging, 22761 Hamburg, Germany}
\affiliation{University of Potsdam, Institute of Physics and Astronomy, 14476 Potsdam, Germany}

\author{Zeger Hens}
\email{zeger.hens@ugent.be}
\affiliation{Physics and Chemistry of Nanostructures, Ghent University, Gent 9000, Belgium}
\affiliation{NoLIMITS Center For Non-Linear Microscopy and Spectroscopy, Ghent University, Gent, Belgium}

\author{Kartik Ayyer}
\email{kartik.ayyer@mpsd.mpg.de}
\affiliation{Max Planck Institute for the Structure and Dynamics of Matter, 22761 Hamburg, Germany}
\affiliation{The Hamburg Center for Ultrafast Imaging, 22761 Hamburg, Germany}


\begin{abstract}
The outstanding opto-electronic properties of lead halide perovskites have been related to the formation of polarons. Nevertheless, the observation of the atomistic deformation brought about by one electron-hole pair in these materials has remained elusive. Here, we measure the diffraction patterns of single $\mathrm{CsPbBr}_3$ quantum dots (QDs) with and without resonant excitation in the single exciton limit using serial femtosecond crystallography (SFX). By reconstructing the 3D differential diffraction pattern, we observe small shifts of the Bragg peaks indicative of a crystal-wide deformation field. Building on DFT calculations, we show that these shifts are consistent with the lattice distortion induced by a delocalized electron and a localized hole, forming a mixed large/small exciton polaron. This result creates a clear picture of the polaronic deformation in $\mathrm{CsPbBr}_3$ QDs, highlights the exceptional sensitivity of SFX to lattice distortions in few-nanometer crystallites, and establishes an experimental platform for future studies of electron-lattice interactions.
\end{abstract}

\maketitle

A polaron is a quasi-particle in a crystalline lattice consisting of a material object, such as an electron, and an accompanying lattice deformation field. Large polarons have deformation fields extending well beyond a single unit cell, while small polarons involve a more localized lattice distortion~\cite{Franchini:2021}. In particular in the case of lead halide perovksites (LHPs), which developed in recent years from an absorber material in highly efficient solar cells to a multipurpose semiconductor for detecting and emitting light~\cite{Shamsi:2019,Eperon:2017}, experimental and computational studies have related specific opto-electronic characteristics to polaron formation~\cite{Ghosh:2020}. Large polarons, for example, have been linked to the enhanced charge-carrier lifetime,\cite{Lan:2019} long diffusion length~\cite{Zhu:2015}, and slow second-order electron-hole recombination~\cite{deQuilettes:2019}. However, as many studies on polarons in LHPs rely on computational methods for their interpretation, the need remains for experimental verification of the role polarons play in charge transport in LHPs~\cite{Franchini:2021}; a task hampered by the lack of a direct observation of the polaron-related lattice distortion.

\begin{figure*}[tb]
 \centering
 \includegraphics[width=\textwidth]{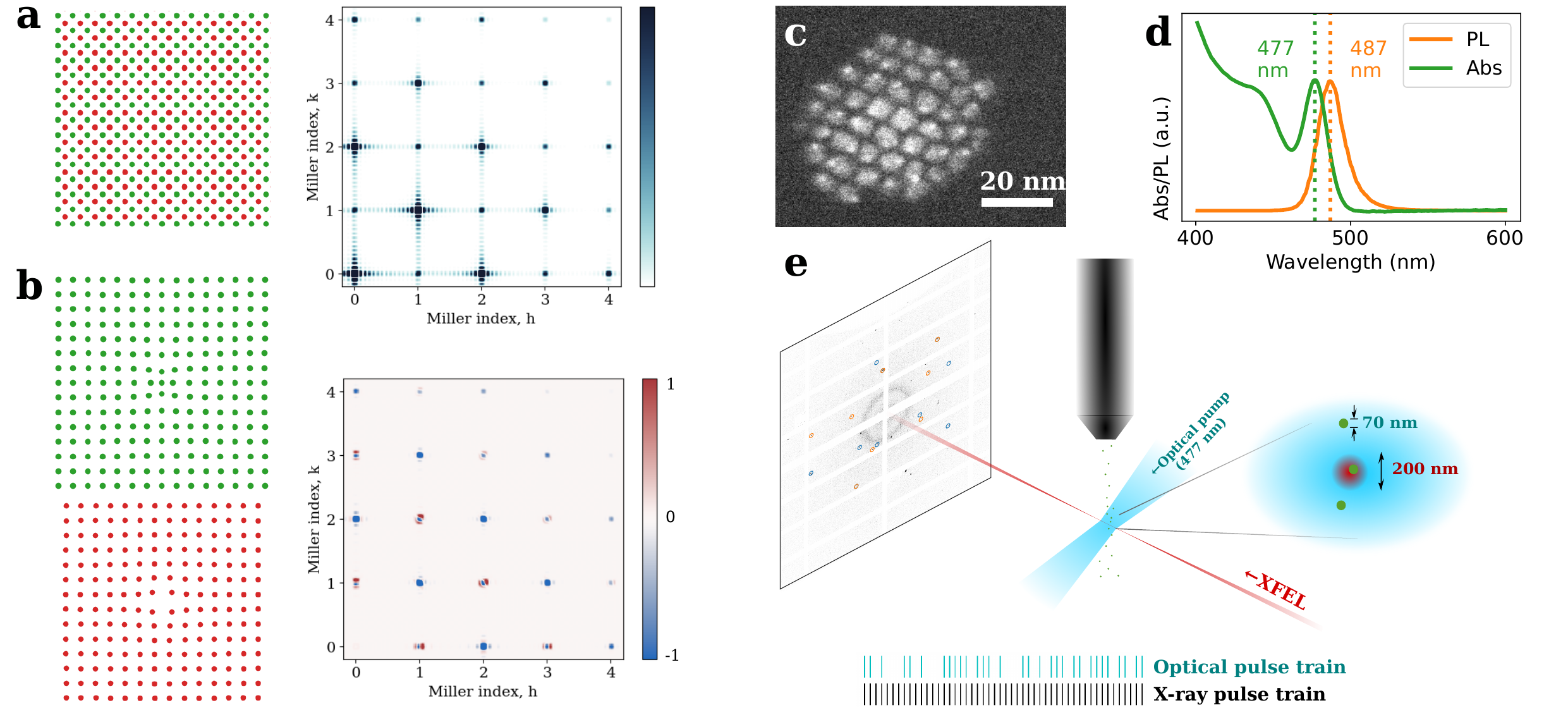}
 \caption{\textbf{Toy model and experimental schematic}. \textbf{a}. Toy model. Illustration of a diatomic body-centred cubic NC with the associated 3D diffraction map in the $hk$ plane. \textbf{b}, Lattice distortion of the two sub-lattices and the resultant differential diffraction map (positive differences in red).\textbf{c}, Representative transmission electron microscope image and \textbf{d}, Absorbance and photoluminescence (PL) spectra of the $\mathrm{CsPbBr}_3$ NCs used in this study. \textbf{e}, Experimental setup. Aerosolized particles are intercepted by the XFEL beam to produce diffraction patterns on the detector. Half the pulses are preceded by an optical pump pulse with a pump-probe delay of 3 ps. The pattern of pulses with an excitation within the first 50 pulses of the European XFEL pulse train is illustrated at the bottom. The inset shows the interaction region with a 200 nm XFEL focus intercepting $\sim70$ nm droplets consisting of the NCs and non-volatile buffer components. }
 \label{fig:experiment}
\end{figure*}

Over the last 5 years, several studies have investigated changes in the atomic lattice of LHP nanocrystals (NCs) after optical pumping with femtosecond (fs) or picosecond (ps) time resolution using pulsed x-ray or electron probes. For such studies, NCs have the advantage of hosting well-defined optical excitations, such as strongly confined electron-hole pairs or bound two-dimensional excitons~\cite{Gramlich:2022,Geiregat:2024}. Using 80 ps X-ray pulses, for example, 11.2~nm $\mathrm{CsPbBr}_3$ NCs were shown to undergo heating-induced phase transitions~\cite{Kirschner:2019}, while 2D LHPs exhibited an anisotropic lattice expansion~\cite{Cuthriell:2022}. Furthermore, a rapid, sub-ps buildup of lattice distortions was observed on 10 nm $\mathrm{CsPbBr}_3$ NCs by means of femtosecond electron diffraction~\cite{Seiler:2023}, while the same method was used to estimate the electron-phonon coupling strength in these materials~\cite{Yazdani:2024}. However, while showing the potential of optical pump/diffractive probe methods to analyse photo-induced changes in the crystal structure, these studies invariably used non-resonant optical pumping at power densities that created multiple excitations per NC, and which analysed azimuthally-averaged diffraction profiles. To observe the atomic lattice distortion caused by a single electron-hole pair -- the key characteristic of an exciton-polaron in such systems -- both excess heat and polarisation-field overlap must be avoided. Such conditions require resonant excitation at power densities that create only a single excitation per NC, presumably in combination with a 2D or 3D reciprocal space map of the light-on/light-off differential diffraction. 

In this study, we use the light-on/light-off diffraction difference of femtosecond X-ray pulses generated by a free electron laser (XFEL) to determine the deformation field in 4.9~nm cubic $\mathrm{CsPbBr}_3$ NCs after resonant excitation. XFELs have been used to study deviations from crystalline order at ultrafast timescales either on single crystals~\cite{Trigo:2013,Wall:2018} or powders~\cite{Abbey:2016,Mariette:2021}, including on LHP single crystals~\cite{Guzelturk:2021}. However, to be sensitive to the small deformation field of single exciton-polarons, we moved from analyzing NC powders to serial femtosecond crystallography (SFX). In SFX, one snapshot at a time is taken on a series of single NCs with and without photo-excitation, for which we used a fixed pump/probe delay. Inspired by structural studies of small-molecule systems~\cite{Schriber:2022,Beyerlein:2015} and ultrafast dynamics of proteins~\cite{Aquila:2012,Barends:2015,Pande:2016}, we reconstructed the 3D diffraction pattern in reciprocal space by indexing the observed Bragg peaks for each NC to determine the crystal orientation~\cite{Chapman:2011}. When obtained from a probe that is coherent across the entire crystal, such a 3D diffraction map is exquisitely sensitive to lattice distortions. Picometer sensitivities have been reported, for example, using the Bragg coherent diffractive imaging technique~\cite{Pfeifer:2006}. The principle is illustrated in Figures~\ref{fig:experiment}a-b. Here, red and green dots represent atomic columns in real space of a small cubic NC of a fictitious diatomic compound with a body-centred structure. Fig.~\ref{fig:experiment}a shows a NC as cut from the bulk, and the corresponding diffraction map sliced normal to the $\langle 001\rangle$ axis. Each Bragg peak is convolved with the so-called \emph{shape transform}, which is the Fourier transform of a 3D mask that is 1 inside and 0 outside the NC. In Fig.~\ref{fig:experiment}b, a radial deformation is added to the NC, which is different in direction for the two sub-lattices. This deformation field changes the diffraction map, which leads to a differential diffraction for each Bragg peak as shown in  Figure~\ref{fig:experiment}b.

\begin{figure*}
 \centering
 \includegraphics[width=\textwidth]{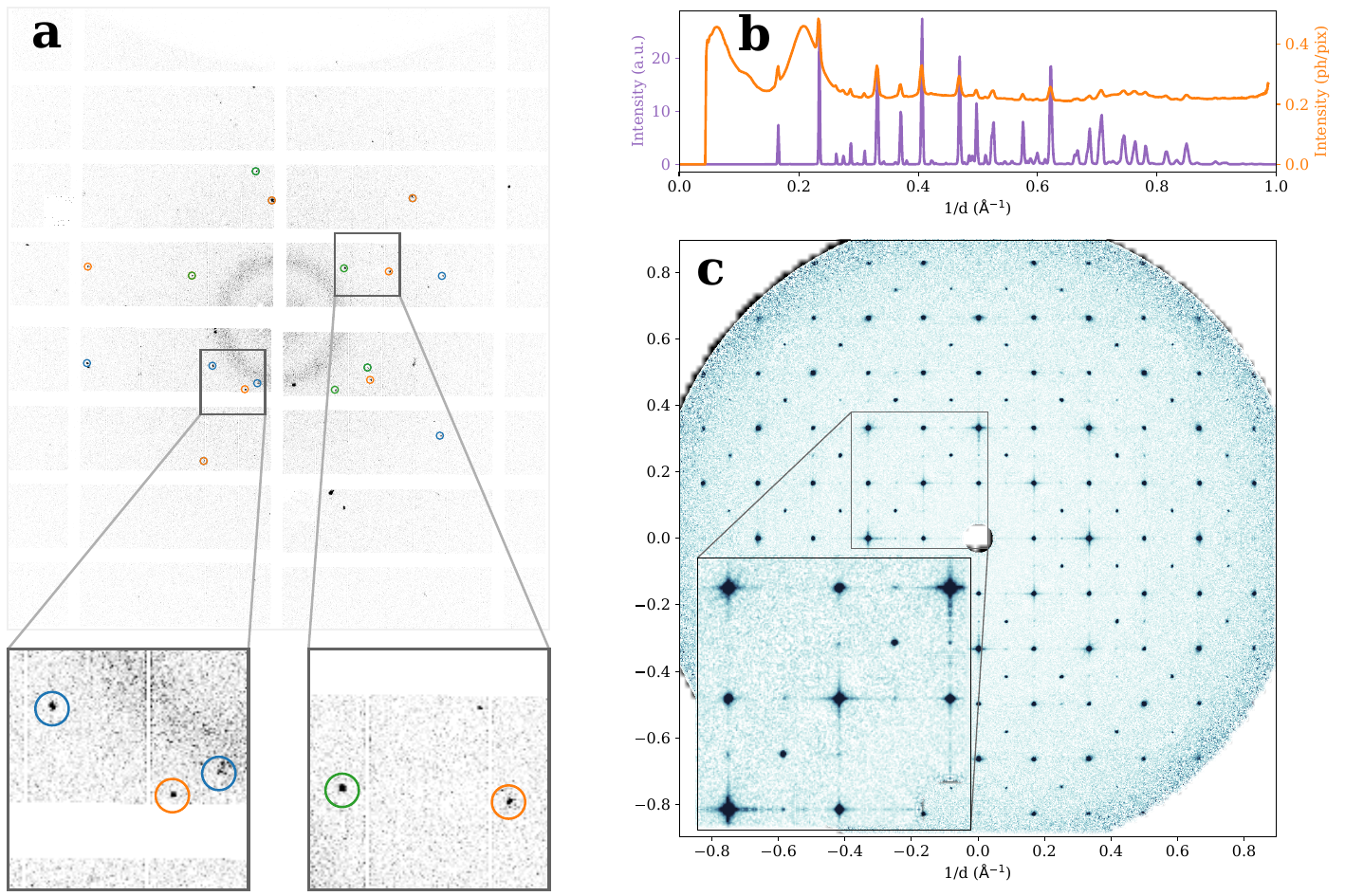}
 \caption{\textbf{Diffraction data and reconstruction.} \textbf{a}, An example of a 3-crystal indexed pattern. The peaks of each crystal are circled by different colours. The diffuse ring corresponds to the signal from the non-volatile components of the buffer such as the ionic liquid. \textbf{b}, Virtual powder plot in purple calculated by integrating Bragg peak intensities after peak finding from individual patterns. Note that by only including the signal from peaks, one avoids contributions from diffuse background and obtains peaks sharper than the peak-width on the detector. The azimuthally averaged intensity calculated directly from the detector frames, analogous to a conventional powder diffraction measurement is shown in orange. One can observe the diffuse background due to the non-volatile solvent such as the ionic liquid and the relative suppression of high angle peaks. \textbf{c}, The central slice ($q_x=0$) of the average \emph{dark} (unpumped) intensity obtained by merging whole patterns according to the orientations predicted by indexing the peaks in 44000 patterns after appropriate diffuse background subtraction. The inset shows a sub-region with an expanded colour scale to visualise the peak shapes and tails.}
 \label{fig:expdata}
\end{figure*}

\section*{Results}
\subsection*{Time-resolved aerosol serial femtosecond crystallography}
We synthesized a batch of $\mathrm{CsPbBr}_3$ NCs using previously published protocols based on size-selective precipitation~\cite{Maes:2018}. As evidenced by the dark field scanning transmission electron microscopy image and the absorption spectrum shown in Figure~\ref{fig:experiment}c-d, these NCs have an average diameter of 4.9 nm and exhibit an exciton transition at 477 nm. Since the shift of this transition with respect to bulk $\mathrm{CsPbBr}_3$ has been assigned to partial confinement of charge carriers~\cite{Geiregat:2024}, we will refer to these NCs henceforth as quantum dots (QDs). In line with previous studies~\cite{Brennan:2017}, the exciton emission has a maximum intensity at 487 nm, which corresponds to a 53 meV Stokes shift. As outlined in Fig.~\ref{fig:experiment}e, we exposed these QDs a few at a time to an XFEL pulse by means of an aerosol sample delivery system that was originally developed for imaging single biomolecules~\cite{Bielecki:2019}. Upon aerosolization, the non-volatile components of the solvent mixture formed a 70 nm droplet which contained on average 1.6 QDs each. The XFEL pulse is strong enough to destroy the sample in a single pulse. Through this delivery method, the sample is refreshed after each exposure while avoiding the strong background scattering produced by liquid jets commonly used for SFX on protein crystals. To the best of our knowledge, a similar approach has only been reported for NCs with dimensions of a few 100 nm~\cite{Chapman:2011,Awel:2018}, and for some fibre crystals with diameters of tens of nm, or longer~\cite{Rodriguez:2015,Wojtas:2017}.

\begin{figure*}
 \centering
 \includegraphics[width=\textwidth]{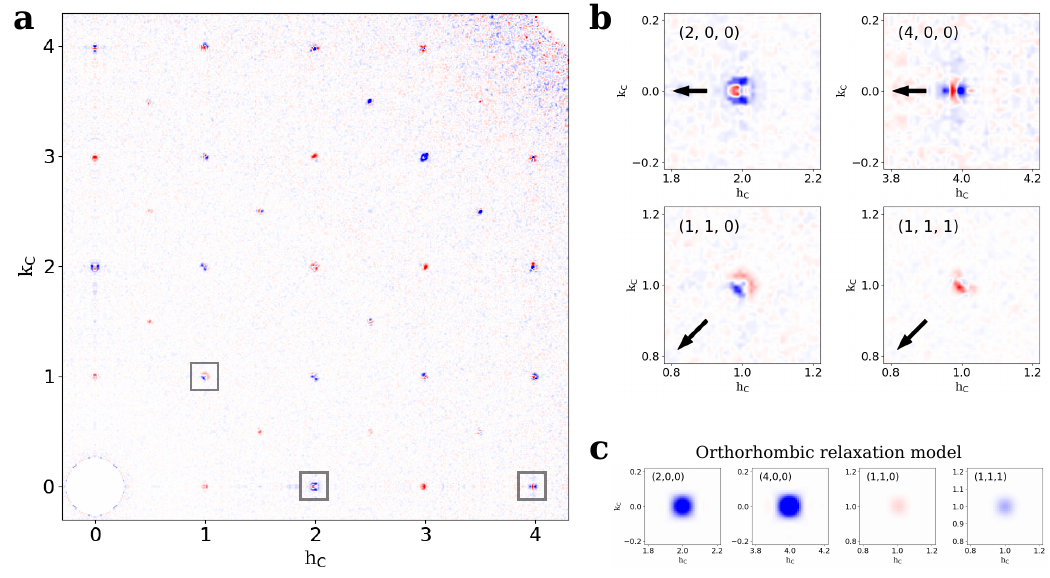}
 \caption{\textbf{Difference intensities upon optical excitation.} \textbf{a}, Slice through the $l_C=0$ plane of the merged intensity difference before and 3 ps after optical excitation. Values in red indicating regions with excess intensity after optical pumping. \textbf{b}, Expanded view of four selected Bragg peaks (arrows point towards $\mathbf{q}=0$), three of which are highlighted in \textbf{a}. For the (111) peak, $l_C=1$. \textbf{c}, Predicted intensity differences from a commonly applied orthorhombic relaxation model where the octahedral tilt varies across the crystal. Note how this distortion does not result in peak shifts, but just changes in integrated intensities.}
 \label{fig:peakshape}
\end{figure*}

Figure~\ref{fig:expdata}a represents a typical diffraction snapshot containing Bragg spots recorded as part of a sequence of 77182 frames (detailed statistics in Supplemental Information S1). In this particular frame, we identified diffraction from three different QDs, as indicated by the coloured circles. Throughout the experiment, half the frames were collected 3 ps after resonant excitation of the QDs, which is long enough for any lattice distortion to settle\cite{Yazdani:2024}, and short enough to prevent heat generation from biexciton recombination.\cite{Geiregat:2024} Using a 120 fs optical pulse with a wavelength of 477 nm and a fluence of 67.5~$\mathrm{\mu}$J/cm$^2$, we expect to probe one electron-hole pair per QD in $\approx 75\%$ of the pumped frames, see Supplemental Information S2. Unpumped and pumped snapshots were examined for Bragg peaks and considered a \emph{hit} when containing at least two identified peaks. For each hit, Bragg peaks were indexed in the orthorhombic \emph{Pnma} space group, and assigned to one or more QDs as illustrated by the coloured circles in Fig.~\ref{fig:expdata}a (see Methods for details). In total, 31547 crystals were indexed from 20103 frames with detected peaks.

From the indexed snapshots, the azimuthally averaged \emph{virtual powder} pattern can be recovered in two ways, as illustrated in Fig.~\ref{fig:expdata}b. The orange trace is the azimuthal average $\langle \bar{I}(q) \rangle$ of all snapshots, where $q$ is the magnitude of the reciprocal wave vector $\bm{q}=(q_x,q_y,q_z)$, expressed as the reciprocal of the $d$-spacing. This intensity corresponds to the quantity that would be obtained from conventional powder diffraction. The purple trace, on the other hand, represents the intensity $\bar{I}(q)$ of the Bragg peaks, averaged over all hits after single-shot peak finding. Clearly, the serial recording of diffraction snapshots of single QDs strongly enhances the measurement sensitivity. In particular the diffuse background at all $q$, and a broad but pronounced feature at around 0.2 \AA$^{-1}$ are almost entirely removed in $\bar{I}(q)$. We separately merged the integrated peak intensities from the diffraction snapshots recorded without and with optical pumping, and used the resulting patterns $\bar{I}_{dark}(q)$ and $\bar{I}_{pump}(q)$ to solve for the average unit cell using the SHELXL software to $1.1\,$\AA$\,$ resolution~\cite{Sheldrick:2015}. However, no statistically significant difference between the two patterns was observed in either the integrated peak intensities, the virtual powder patterns or the azimuthal average intensity. This observation agrees with a previous report, where no change in the electron diffraction powder pattern was observed on 9.5 nm $\mathrm{CsPbBr}_3$ NCs excited using 400 nm pulses with a fluence of 800~$\mathrm{\mu}$J/cm$^2$, 12-fold of what was used in the experiments reported here~\cite{Yazdani:2024}.

More interestingly, having an indexed series of diffraction snapshots of single QDs -- and thus the QD orientation -- enabled us to move beyond azimuthal averaging and reconstruct an average diffraction pattern $\bar{I}(\bm{q})$ in 3D reciprocal space. To do so, we masked the region around Bragg peaks from other QDs in the same detector frame and subtracted a scaled diffuse background for each indexed QD before merging all pixels (see Methods for details). Figure~\ref{fig:expdata}c displays a central slice of the resulting \emph{dark} pattern normal to the $q_x$-axis. Various features can be qualitatively identified in this intensity distribution. The square grid made up by the bright Bragg peaks is characteristic of an approximately cubic lattice with an average lattice constant of $5.90\,$\AA, while additional weak peaks can be observed at half-integer positions in line with previous orthorhombic crystal structure assignments~\cite{Cottingham:2016}. In the rest of the article, we will use the effective cubic lattice peaks with Miller indices $(h_\mathrm{C}, k_\mathrm{C},l_\mathrm{C})$. The bright peaks also exhibit lattice truncation streaks along the $\langle 100 \rangle$ directions due to the approximately cubic shape of individual QDs. Finally, the intensity distribution of the Bragg peaks is somewhat asymmetric around the reciprocal lattice point, which probably reflects inherent strain in these QDs even before optical excitation. We do not observe fringe contrast along the $\langle 100 \rangle$ streaks because the formation of the merged pattern involves averaging over QDs with slightly different sizes and inherent strain.

\subsection*{Optically induced lattice deformations}

Figure~\ref{fig:peakshape}a displays a slice of the diffraction intensity difference in the $q_x-q_y$ plane, $\Delta\bar{I}(q_x,q_y,q_z=0)$, obtained by subtracting $\bar{I}_{dark}$ from $\bar{I}_{pump}$. Interestingly, while the virtual powder patterns $\bar{I}_{pump}$ and $\bar{I}_{dark}$ were similar, Bragg peaks in this differential diffraction pattern show an intricate combination of enhanced (red) and reduced (blue) intensity as a result of optical pumping. Especially for the (200) and (110) diffraction, these patterns can be interpreted in first approximation as an inward or outward shift of the Bragg peak, respectively. Given the direct link between the diffraction pattern and the atomic structure, we thus conclude that the resonant formation of a single electron-hole pair in 4.9 nm $\mathrm{CsPbBr}_3$ QDs comes with a distortion of the atomic lattice, i.e., the formation of an exciton-polaron. 

Qualitative observations of the differential diffraction provide a first understanding of the lattice deformation field. First, difference scattering signals in Fig.~\ref{fig:peakshape}a are concentrated near the Brillouin zone (BZ) centre, indicative of long-range correlations in the deformation field. A strongly localized lattice distortion will produce difference scattering throughout reciprocal space, but this is not visible in the data. In fact, outside of the crystal truncation streaks resulting from the cubic shape of the NCs, no diffuse scattering is observed far from the BZ centre. Second, the differential diffraction appears to be primarily in the radial direction and not along, for example, the $\langle 100\rangle$ facets. This finding is highlighted by the differential diffraction of four representative peaks in Fig.~\ref{fig:peakshape}b, where the arrow represents the direction towards the origin. Such systematics suggest a predominantly spherically symmetric deformation field, although the presence of off-axis or tangential features means that the deformation field is not perfectly isotropic. 

A commonly proposed deformation field involves a relaxation of the orthorhombic distortion of the Br octahedra upon optical excitation~\cite{Miyata:2017,Seiler:2023,Yazdani:2024}, see Supplementary Information S3. Figure~\ref{fig:peakshape}c displays the differential diffraction such a distortion would cause (see Methods for details). Interestingly, this deformation field yields a differential diffraction that mostly reflects small changes in the integrated diffraction intensity, rather than the experimentally observed shifts of the Bragg peaks. It thus appears that a single exciton-polaron created through resonant excitation leads to a different lattice distortion than multiple electron-hole pairs formed by non-resonant excitation.

\begin{figure*}
 \centering
 \includegraphics[width=\textwidth]{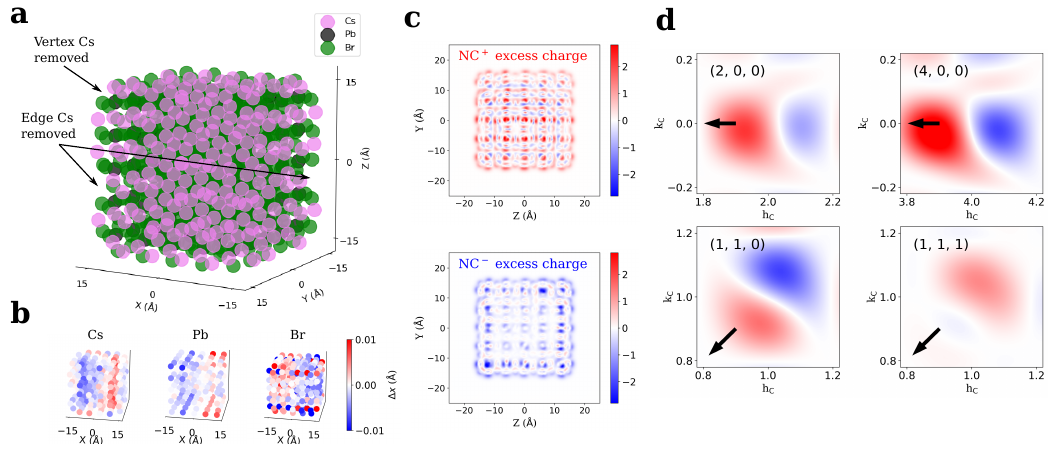}
 \caption{\textbf{Density functional theory (DFT) modelling}. \textbf{a}, Representation of one of the constructed 3 nm QDs. Some of the 16 Cs atoms removed for charge neutrality are indicated. \textbf{b}, Atomic displacements along the $x$ direction for the different sub-lattices between the relaxed triplet and singlet states. Other than the outermost layer, one can visualize the outward displacements of the Cs and Pb atoms and inward displacement of the Br atoms. \textbf{c}, Excess charge density maps for the cationic and anionic NCs respectively, projected along the $x$ axis, showing the localization of the hole and excess negative charge concentrated in the outer regions. \textbf{d}, Predicted intensity difference for the same peaks as in Fig.~\ref{fig:peakshape}b. Note the broader peaks due to the 3~nm simulated particle.}
 \label{fig:dft}
\end{figure*}

\subsection*{Density functional theory modelling}

A 3D diffraction difference map is a rich source of information that encodes the deformation field $\bm{u}$ in reciprocal space. However, the differential diffraction measured here results from subtracting light-on/light-off diffraction intensities calculated as separate averages over an ensemble of differently sized QDs, and cannot be used for the \emph{ab-initio} reconstruction of the deformation field. To create a first benchmark relating the experimental differential diffraction to the deformation field of the exciton-polaron in $\mathrm{CsPbBr}_3$ QDs, we calculated the atomistic structure of $\mathrm{CsPbBr}_3$ QDs using density functional theory (DFT). As outlined in the Methods section, we relaxed the geometry of $\approx 3\:\mathrm{nm}$ $\mathrm{CsPbBr}_3$ QDs with the brute formula $\mathrm{Cs}_{200}\mathrm{Pb}_{125}\mathrm{Br}_{450}$ in the $S=0$ singlet ground state and the $S=1$ triplet excited state. While the latter yields the triplet state of the exciton, we assumed that a spin flip of the exciton has no significant impact on the atomic geometry. For both states, we obtained the coordinates of the different atoms and the density of the valence electrons. From these data, we calculated diffraction intensity differences that can be compared directly with the corresponding slices of the experimental 3D differential diffraction. A full overview of the DFT approach and results is provided in Supplementary Information S4.

Figure~\ref{fig:dft}a represents the relaxed structure of one of the $\mathrm{CsPbBr}_3$ QD models, which are cut as cubes from the bulk $\mathrm{CsPbBr}_3$ lattice. To ensure charge neutrality, we removed the 8 vertex Cs atoms and 8 additional edge Cs atoms. Five models were considered, for which different edge Cs atoms were removed. Averaged over these 5 models, we obtained a mean relaxation energy when optimizing the geometry of the $S=1$ triplet excited state starting from the $S=0$ singlet ground state of $56$~meV. This number would correspond to a Stokes shift of $112$~meV, which exceeds the experimentally observed shift of $53$~meV for 4.9 nm $\mathrm{CsPbBr}_3$ QDs, but is comparable to extrapolated shifts for 3 nm QDs~\cite{Brennan:2017}. 

Next, we obtained $\bm{u}$ from the difference in relaxed atomic positions of the $S=1$ excited and the $S=0$ ground state. Figure~\ref{fig:dft}b shows $u_x$ -- the $x$-component of the displacement -- of the three atomic sub-lattices for the $\mathrm{CsPbBr}_3$ QD model shown in Fig.~\ref{fig:dft}a. Here, atoms are coloured red or blue depending on $u_x$ being positive or negative. While the relaxation is fairly complex, this colour coding underscores that, on average, Cs and Pb cations move outwards -- positive shift at the right side and negative shift at the left side of the QD -- while Br anions move inwards. Combining all field components yields, to first approximation, a radial deformation field with outward shifts for the cations and inward shifts for the anions (see Methods for details). Such a longitudinal field is consistent with a charge distribution that involves a central positive charge and a distributed negative charge. This conclusion is illustrated by the projected excess charge maps obtained by removing or adding an electron to the $\mathrm{CsPbBr}_3$ QD, see Fig.~\ref{fig:dft}c, and is consistent with recent spectroscopic evidence obtained on the same system~\cite{Cannelli:2021,Geiregat:2024}. Furthermore, the differential diffraction calculated from the $S=1$ excited and the $S=0$ ground state yields a pattern of shifted Bragg peaks that point towards the centre of reciprocal space, see Fig.~\ref{fig:dft}d. Even so, while calculated and experimental differential diffraction are consistent around the 200 Bragg peak, they are not around 110 and 111. For those peaks, the predicted  inward shift (Fig.~\ref{fig:dft}d) contrasts with a measured outward shift (Fig.~\ref{fig:peakshape}b). We thus conclude that DFT calculations on 3.0 nm $\mathrm{CsPbBr}_3$ QDs do not fully grasp the average structure of the exciton-polaron in 4.9 nm $\mathrm{CsPbBr}_3$ QDs.

\begin{figure*}
 \centering
 \includegraphics[width=\textwidth]{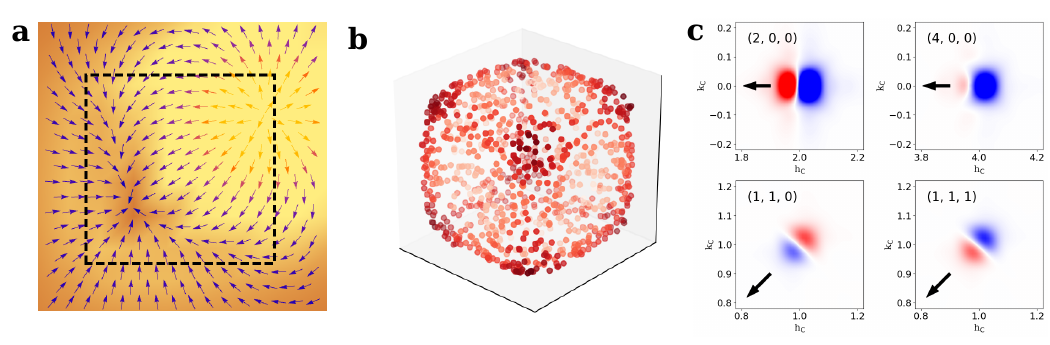}
 \caption{\textbf{Random hole localization.} \textbf{a}, Electric field distribution in the central plane resulting from a localized hole near the surface and a delocalized electron density. The dashed line represents the boundary of the NC. \textbf{b}, Sampled hole positions in the volume of the $\mathrm{CsPbBr}_3$ NC (color represents the distance from the centre). \textbf{c}, Predicted average intensity difference for the same peaks as in Fig.~\ref{fig:peakshape}b calculated by averaging the sampled hole positions in \textbf{b}.}
 \label{fig:randloc}
\end{figure*}

\subsection*{Random hole localization model}

The DFT calculations yielded a mostly radial deformation field $\bm{u}$, where cations shifted outward and anions inward in agreement with the electric field of a charge distribution involving a localized positive and a delocalized negative charge. We took such a presumed exciton charge distribution as a starting point to get an improved estimate of $\bm{u}$ from the balance between the electric force and the restoring elastic force on each atom. Considering the restoring force constant as an adjustable parameter, we tuned the relative displacement of the different atoms and compared predicted and experimental diffraction differences. However, no matter the ratio of the relative displacement amplitudes, a central positive charge never yielded an outward shift of the 110 peaks using this approach (see Supplemental Information S5). We therefore concluded that the experimental differential diffraction is not the mere average of exciton-polarons characterized by a central positive charge over an ensemble of photo-excited QDs.  

As can be seen in Fig.~\ref{fig:dft}c, removing one electron from a $\mathrm{CsPbBr}_3$ QD yields an excess positive charge that is somewhat off-centre, and has a considerable surface contribution. We conjectured that this deviation from a central charge will be more pronounced for the larger QDs that we analysed experimentally. Hence, to better describe the average deformation field of many QDs, we calculated the diffraction from an ensemble of pumped QDs with hole charges randomly localized within the QD. The displacement of each atom is then proportional to the net field from a point positive charge and a delocalized negative charge, see Fig.~\ref{fig:randloc}a. This approach consistently yielded inward shifts of the 200 Bragg peak, in agreement with the experimental differential diffraction. Furthermore, by introducing a bias that favours hole localization closer to the QD surface (illustrated in Fig.~\ref{fig:randloc}b), we can reproduce the experimentally observed combination of an inward shift of the 200 and an outward shift of the 110 Bragg peak, see Fig.~\ref{fig:randloc}c. This approach also captures the main characteristics of the 400 and 111 Bragg peaks. By scanning the relative displacement field amplitudes (see Supplemental Information S3), we also observe that the correct peak shifts for the 200 peak requires that the Br atoms are more weakly restrained than the Cs and Pb atoms. This result indicates that upon photo-excitation, an exciton-polaron is formed that consists of a delocalized electron and a localized hole, the position of which is biased towards the outer parts of the QD. This charge distribution induces atomic displacements proportional to the local electric field and the ionic charge, with Br atoms displacing more than the Cs and Pb atoms.

\section*{Discussion}

By means of serial femtosecond crystallography, we reconstructed the 3D diffraction pattern in reciprocal space of 4.9 nm $\mathrm{CsPbBr}_3$ QDs. Comparing light-on/light-off diffraction, we obtained differential diffraction patterns that featured subtle changes after photo-excitation that are not resolved in the azimuthally averaged powder pattern. More precisely, the 3D differential diffraction map in reciprocal space shows a radial pattern that reflects, most simply, inward or outward shifts of the Bragg peaks. Supported by density functional theory calculations, we argue that in real space, these shifts reflect the formation of an exciton-polaron, which distorts the lattice by displacing cations outward and anions inward. This distortion resembles a longitudinal deformation brought about by the electric field of a localized positive and a delocalized negative charge. A more detailed analysis indicates that the opposite shifts of the 200 and 110 Bragg peaks result from the ensemble averaging -- intrinsic to the SFX approach -- of exciton-polarons having a different, surface-biased hole localization. Interestingly, in agreement with the partial confinement of charge carriers in $\mathrm{CsPbBr}_3$ QDs~\cite{Geiregat:2024}, this result puts forward the exciton-polaron in $\mathrm{CsPbBr}_3$ QDs as a mix between a large and a small polaron.  

Our study provides the first structural evidence that resonant excitation in the single exciton limit leads to the formation of exciton-polarons in $\mathrm{CsPbBr}_3$ quantum dots. By highlighting the unmatched sensitivity of SFX to small reorganisations of the atomic lattice, such as the strain profile associated with a single exciton-polaron, this results creates a vast range of opportunities for researchers to investigate the interaction between free charges, excitons or excitonic complexes and the atomic lattice. Given the femtosecond time resolution of the approach, one can envision monitoring the dynamics of polaron formation and relaxation in reciprocal space by sweeping the pump-probe time delay. In future experiments, the serial approach can also be used to classify snapshots by the size and shape of the QDs, creating a subset of frames that can be studied quantitatively~\cite{Ayyer:2021,Zhuang:2022}. For a perfectly homogeneous ensemble, the 3D diffraction pattern is directly related to the Fourier transform of the electron density in the QDs~\cite{Pfeifer:2006}. In that way, the deformation field accompanying a polaron can be unambiguously identified~\cite{Clark:2013}, a major step beyond what ensemble-averaging powder diffraction approaches can achieve. 

\section*{Methods}
\subsection*{Pump-probe aerosol serial femtosecond crystallography}
Measurements were performed in the upstream interaction chamber of the Single Particles Biomolecules and Clusters/Serial Femtosecond Crystallography (SPB/SFX) instrument at the European XFEL. 13.6~keV X-ray photons were focussed to a 200-nm spot using Kirkpatrick-Baez mirrors. The Adaptive Gain Integrating Pixel Detector (AGIPD) was placed 122.4~mm downstream of the interaction region to collect the diffraction from the aerosolized particles. Diffraction patterns were collected at a rate of 3520~frames/s with approximately 1\% of the frames containing measurable diffraction from nanocrystals.

In order to deliver the samples to the X-ray beam, nanocrystals were aerosolized using electrospray ionization and then transported and focussed to the X-ray beam using an aerodynamic lens stack~\cite{Bielecki:2019}. Electrospray ionization requires a conductive solvent in order to obtain the Taylor cone necessary for the formation of small  droplets. To achieve the aerosolization of the $\mathrm{CsPbBr}_3$ NCs which do not survive in polar solvents, they were dispersed in toluene and 2\% ionic liquid (trihexyltetradecylphosphonium bis(trifluoromethylsulfonyl)amide) was added to make the solvent conductive. Prior tests showed that the QDs survived in this environment for at least a few hours before the NCs dissolved, as evidenced by the formation of lead bromide precipitates and the dispersion not fluorescing.

\subsection*{Data reduction and recalibration}
Peak finding to detect patterns with crystalline diffraction was performed with the OnDA implementation~\cite{Mariani:2016} of the \emph{peakfinder8} algorithm~\cite{Barty:2014}. The crystal hits were indexed using a modified version of the \emph{SPIND} algorithm~\cite{Li:2019}. Instead of pre-generating a large reference table which consists of all possible pairs of ideal Bragg peaks as in the original \emph{SPIND} implementation, we leveraged the fact that the $|\mathbf{q}|$ of Bragg peaks in reciprocal space are orientation independent and generated the reference pairs on the fly. Multiple crystal indexing was done by removing the previously indexed peaks in each step and attempting indexing again. To increase the indexing success rate, we adopted an adaptive criterion for accepting a set of indexed peaks as a crystal. In the first round, we set the criterion to find at least four peaks that fit a single lattice. In the second round, for patterns where we could not find any crystal, we lowered the criterion to 3 peaks. 

The orientations of the indexed crystals, denoted as a matrix $\Omega$, were further refined by explicitly modelling the cross shape of the Bragg peaks, resulting from the cubic shape of the NCs. Specifically, we minimised the following target function against $\Omega$:
\begin{equation}
    \sum_{\mathbf{q}, s \in \text{Crystal}} s \cdot \exp\biggl(
      - \frac{1}{2 W^2}\Bigl(|\Omega \mathbf{q}|^2 - r \cdot \max_{i \in \{x, y z\}} (\Omega\mathbf{q})_i^2\Bigr)
    \biggr),
\end{equation}
where $s$ is the integrated peak intensity. The hyper parameters were the peak width $W = 4.6\,\text{voxel}$ and the bias ratio $r=0.3$.

Since we were interested in both the strong peak intensities as well as the weak tails of the peaks, the AGIPD was operated in the gain-switching mode. The patterns containing peaks were recalibrated using a custom procedure which used the data from the in-pixel constant current source~\cite{Allahgholi:2019} to better calibrate pixels in the regions where the \emph{gain} signal of the AGIPD indicated the pixel was close to, or had crossed, the gain switching threshold.

\subsection*{3D intensity generation}
The orientations from indexing were used to rotate the Ewald sphere for the entire frame and merge it into the 3D reciprocal space. To account for the effects of beam fluence, sample crystal size and partiality of Bragg peaks, the indexed crystals of each frame were independently rescaled using different rescaling factors. The rescaling process involved first merging the intensities from all the indexed crystals together to create a 3D intensity reference without any rescaling. Then, for each crystal, the corresponding Ewald sphere cut of the reference intensity was extracted from the 3D intensity map, and a rescaling factor for that crystal was determined by calculating the ratio between the mean intensity of the brightest 10 pixels in the reference cut and the corresponding crystal's mean intensity. 


To refine the intensity reconstruction, we implemented a process to remove crystals that deviated significantly from their corresponding merged intensity slice. For each indexed crystal, following normalization to the brightest pixel, a selected set of pixels of the pattern were compared to the merged intensities. These pixels were within a radius of 4 pixels from expected the Bragg peaks, which had expected intensities greater than 0.01 times the mean intensity. The scalar product of these pixels with the corresponding slice of the 3D merged intensity was used as a metric to reject indexed crystals which were too dissimilar.

Before merging the frames into reciprocal space, the pixels in the neighbourhood of peaks in the pattern, but not indexed into the lattice were excluded. This significantly reduced the artifacts in the merged volume from multi-crystal diffraction patterns. Background subtraction was carried out in two sequential steps for indexed crystal patterns. First, the radial average for each pattern was calculated, excluding outlier values from the vicinity of Bragg peaks, generating a one-dimensional (1D) array depicting the background variation with respect to $|\mathbf{q}|$ for each frame. Following this, Singular Value Decomposition (SVD) was applied on these 1D feature arrays, with the first three components employed to reconstruct the background.

\subsection*{DFT calculation details}
Theoretical calculations were conducted at the Density Functional Theory (DFT) level using the semi-local PBE exchange-correlation functional~\cite{Perdew:1996}, as implemented in the CP2K 2024.1 package~\cite{Kuhne:2020}. A double-zeta basis set, augmented with polarization functions, was employed alongside effective core potentials for all atom types~\cite{Vandevondele:2007}. A standardized protocol was followed: first, a manually selected initial structure was relaxed to its ground state, defined by a spin multiplicity of 1. The resulting optimized ground-state structure was then used as the starting point for relaxing the excited-state structure, defined by a spin multiplicity of 3. For both states, the electron density was subsequently calculated via single-point calculations and exported as cube files, which contain the atomic coordinates and electron density on a 3D grid. These data were then used to determine the diffraction patterns for both the ground and excited states.

The analysis was carried out on a charge-neutral nanocrystal (NC) model with the stoichiometry $\mathrm{Cs}_{200}\mathrm{Pb}_{125}\mathrm{Br}_{450}$. The NC adopts an approximately cubic shape, consisting of 5 cubic unit cells extending in each direction, as illustrated in Fig.~\ref{fig:dft}a. Structurally, this composition results in an inner framework of $5\times5\times5$ Pb atoms, surrounded by an outer framework of $6\times6\times6$ Cs atoms. To achieve a total of 200 Cs atoms, 16 vacancies were introduced into the outer Cs layer. Specifically, all 8 Cs atoms at the vertex positions were removed, and in most cases, the remaining 8 vacancies were positioned along the edges. Different arrangements of these vacancies give rise to distinct model NCs. A comprehensive description of these variations is provided in Supplemental Information S4.

\subsection*{Intensity calculation for different polaron models}
The scattered intensity from a single crystal was calculated by performing a 3D Fourier transform of the electron density of the crystal, represented by point-like atoms weighted by their tabulated scattering factors at 13.6 keV. The simulated intensities for the undistorted crystal were calculated by incoherently averaging the scattered intensities from crystals sampled from a size range between 3 and 15 nm weighted by the Gamma distribution with a shape parameter of 2.0 and an average size of 5.4 nm. The orthorhombic basis reported in Materials Project ID \texttt{mp-567629}~\cite{Jain:2013} was used to place atoms in the unit cell and the crystal was truncated along the $\langle100\rangle$ directions of the cubic unit cell. The shape of the crystal for each size was calculated using a superellipsoid envelope with an exponent of 5.0 in order to partly round the corners and more closely match the observed truncation rods in the measured 3D intensities.

For the orthorhombic relaxation model, the displacement of the atoms from the cubic lattice was parametrized by the linear parameter $t$ such that $t=0$ represented the cubic lattice and $t=1$ represented the tabulated orthorhombic structure. The distortion from the perfect orthorhombic structure for atom $i$ is then
\begin{equation}
\mathbf{\delta}_i = t\,\mathbf{r}_{\mathrm{o},i} + (1-t)\,\mathbf{r}_{\mathrm{c},i}
\end{equation}
where $\mathbf{r}_{\mathrm{o|c},i}$ represents the position of atom $i$ in the orthorhombic or cubic lattice respectively. Supplemental Information S2 shows an illustration of this distortion in real space. The results shown in Fig.~\ref{fig:peakshape}c were obtained by varying the $t$ parameter from 0.8 for the unit cell next to the centre and decaying to 1 with the square of the distance of the unit cell to the origin. This model assumes the octahedral rotation decreases due to the polaron. Assuming an increase of the octahedral rotation close to the centre of the QD results in the opposite sign for intensity changes compared to that displayed in Fig.~\ref{fig:peakshape}c.

For the random hole localization model, the normalized hole positions, $\mathbf{\tilde{r}}_h$, were randomly generated using the following algorithm,
\begin{align}
\mathbf{\tilde{r}}_h &:= (\mathcal{U}(-1,1), \mathcal{U}(-1,1), \mathcal{U}(-1,1))\\
\mathbf{\tilde{r}}_h &:= \mathbf{\tilde{r}}_h \frac{\left\Vert\mathbf{\tilde{r}}_h\right\Vert_5^{0.1}}{\left\Vert\mathbf{\tilde{r}}_h\right\Vert_5}
\end{align}
where $\mathcal{U}(-1,1)$ is a uniform random number between -1 and 1 and $\left\Vert\cdot\right\Vert_5$ refers to the $l^5$-norm of a vector. The actual position was then estimated by scaling this vector by the size of the sampled NC.

For simplicity, the electric field from the delocalized electron charge was calculated from a sphere model:
\begin{equation}
\mathbf{E}(\mathbf{r}) \propto \frac{\mathbf{r}}{(s/2)^3}
\end{equation}
where $s$ is the size of the sampled NC.

\begin{acknowledgments}
\noindent Part of this work was supported by the Cluster of Excellence 'CUI: Advanced Imaging of Matter' of the Deutsche Forschungsgemeinschaft (DFG) - EXC 2056 - project ID 390715994 and by Deutsches Elektronen-Synchrotron (DESY), a member of the Helmholtz Association (HGF). Z.H. acknowledges FWO-Vlaanderen (SBO Proceed, research project G0B2921N) and Ghent University (BOF-GOA 01G02124) for research funding. I.I. acknowledges Horizon Europe EIC Pathfinder program through project 101098649–UNICORN and IKUR Strategy under the collaboration agreement between Ikerbasque Foundation and BCMaterials on behalf of the Department of Education of the Basque Government. Part of the computational resources and services used in this work were provided by the VSC (Flemish Supercomputer Center), funded by the Research Foundation Flanders (FWO) and the Flemish Government – department EWI. We acknowledge European XFEL in Schenefeld, Germany, for provision of X-ray free-electron laser beamtime at SPB/SFX SASE1 under proposal number 2746 and would like to thank the staff for their assistance.  
\end{acknowledgments}

\section*{Author contributions}
\noindent K.A., Z.H., H.L. and H.N.C. conceived the experiment. The SFX experiment was performed by all authors except I.I. For the experiment, samples were prepared by M.S. with the help of O.E. and Z.H., and sample delivery was performed by J.B., A.K.S., A.E., S.R.-Z., L.W. and J.K. Online data analysis during the experiment was performed by K.A., S.Z., Y.Z., T.W. and A.M. The SFX analysis was performed by S.Z. with help from K.A. DFT simulations were performed by Z.H. and I.I. Final modelling and interpretation was led by K.A., S.Z. and Z.H., with input from H.N.C., I.I. and H.L. The manuscript was written by K.A. and Z.H. with input from all authors.

\bibliography{refs}

\end{document}


\title{Supplemental Information: Direct observation of single exciton polarons in CsPbBr$_3$ quantum dots by femtosecond X-ray differential diffraction}

\author{Zhou Shen}
\affiliation{Max Planck Institute for the Structure and Dynamics of Matter, 22761 Hamburg, Germany}

\author{Margarita Samoli}
\affiliation{Physics and Chemistry of Nanostructures, Ghent University, Gent 9000, Belgium}

\author{Onur Erdem}
\affiliation{Physics and Chemistry of Nanostructures, Ghent University, Gent 9000, Belgium}

\author{Johan Bielecki}
\affiliation{European XFEL, 22869 Schenefeld, Germany}

\author{Amit Kumar Samanta}
\affiliation{Center for Free-Electron Laser Science CFEL, Deutsches Elektronen-Synchrotron DESY, 22607 Hamburg, Germany}

\author{Juncheng E}
\affiliation{European XFEL, 22869 Schenefeld, Germany}

\author{Armando Estillore}
\affiliation{Center for Free-Electron Laser Science CFEL, Deutsches Elektronen-Synchrotron DESY, 22607 Hamburg, Germany}

\author{Chan Kim}
\affiliation{European XFEL, 22869 Schenefeld, Germany}

\author{Yoonhee Kim}
\affiliation{European XFEL, 22869 Schenefeld, Germany}

\author{Jayanath Koliyadu}
\affiliation{European XFEL, 22869 Schenefeld, Germany}

\author{Romain Letrun}
\affiliation{European XFEL, 22869 Schenefeld, Germany}

\author{Federico Locardi}
\affiliation{Dipartimento di Chimica e Chimica Industriale, Università degli Studi di Genova, 16146 Genova, Italy}
\affiliation{Physics and Chemistry of Nanostructures, Ghent University, Gent 9000, Belgium}

\author{Jannik Lübke}
\affiliation{Center for Free-Electron Laser Science CFEL, Deutsches Elektronen-Synchrotron DESY, 22607 Hamburg, Germany}

\author{Abhishek Mall}
\affiliation{Max Planck Institute for the Structure and Dynamics of Matter, 22761 Hamburg, Germany}

\author{Diogo Melo}
\affiliation{European XFEL, 22869 Schenefeld, Germany}

\author{Grant Mills}
\affiliation{European XFEL, 22869 Schenefeld, Germany}

\author{Safi Rafie-Zinedine}
\affiliation{European XFEL, 22869 Schenefeld, Germany}

\author{Adam Round}
\affiliation{European XFEL, 22869 Schenefeld, Germany}

\author{Tokushi Sato}
\affiliation{European XFEL, 22869 Schenefeld, Germany}

\author{Raphael de Wijn}
\affiliation{European XFEL, 22869 Schenefeld, Germany}

\author{Tamme Wollweber}
\affiliation{Max Planck Institute for the Structure and Dynamics of Matter, 22761 Hamburg, Germany}
\affiliation{The Hamburg Center for Ultrafast Imaging, 22761 Hamburg, Germany}

\author{Lena Worbs}
\affiliation{Center for Free-Electron Laser Science CFEL, Deutsches Elektronen-Synchrotron DESY, 22607 Hamburg, Germany}

\author{Yulong Zhuang}
\affiliation{Max Planck Institute for the Structure and Dynamics of Matter, 22761 Hamburg, Germany}

\author{Adrian P. Mancuso}
\affiliation{European XFEL, 22869 Schenefeld, Germany}
\affiliation{Department of Chemistry and Physics, La Trobe Institute for Molecular Science, La Trobe University, Melbourne, VIC 3086, Australia}

\author{Richard Bean}
\affiliation{European XFEL, 22869 Schenefeld, Germany}

\author{Henry N. Chapman}
\affiliation{Center for Free-Electron Laser Science CFEL, Deutsches Elektronen-Synchrotron DESY, 22607 Hamburg, Germany}
\affiliation{The Hamburg Center for Ultrafast Imaging, 22761 Hamburg, Germany}
\affiliation{Department of Physics, Universität Hamburg, 22761 Hamburg, Germany}

\author{Jochen Küpper}
\affiliation{Center for Free-Electron Laser Science CFEL, Deutsches Elektronen-Synchrotron DESY, 22607 Hamburg, Germany}
\affiliation{The Hamburg Center for Ultrafast Imaging, 22761 Hamburg, Germany}
\affiliation{Department of Physics, Universität Hamburg, 22761 Hamburg, Germany}

\author{Ivan Infante}
\affiliation{BCMaterials, Basque Center for Materials, Applications, and Nanostructures, UPV/EHU Science Park, Leioa, 48940 Spain}

\author{Holger Lange}
\affiliation{The Hamburg Center for Ultrafast Imaging, 22761 Hamburg, Germany}
\affiliation{University of Potsdam, Institute of Physics and Astronomy, 14476 Potsdam, Germany}

\author{Zeger Hens}
\email{zeger.hens@ugent.be}
\affiliation{Physics and Chemistry of Nanostructures, Ghent University, Gent 9000, Belgium}
\affiliation{NoLIMITS Center For Non-Linear Microscopy and Spectroscopy, Ghent University, Gent, Belgium}

\author{Kartik Ayyer}
\email{kartik.ayyer@mpsd.mpg.de}
\affiliation{Max Planck Institute for the Structure and Dynamics of Matter, 22761 Hamburg, Germany}
\affiliation{The Hamburg Center for Ultrafast Imaging, 22761 Hamburg, Germany}


\maketitle

\section{Data collection statistics}

\begin{table}[h]
    \centering
    \begin{tabular}{l|c|c|c}
         Metric & Dark & Light & Total \\
         \hline 
         Total frames & $17\,582\,224$ & $17\,482\,325$ & $35\,064\,549$\\
         Hit frames & $441\,071$ & $439\,302$ & $880\,373$ \\
         Hit rate & 2.51\% & 2.51\% & 2.51\%\\
         Frames with peaks & 39964 & 37218 & 77182 \\
         Peak hit rate & 0.227\% & 0.213\% & 0.220\% \\
         Indexed frames & 10106 & 9997 & 20103 \\
         Indexed crystals & 15842 & 15705 & 31547\\
         Indexing rate & 39.6\% & 42.2\% & 40.8\%\\
         Indexed peaks & 79855 & 79805 & 159660 \\
    \end{tabular}
    \caption{Data collection statistics for the dark (umpumped) and light (3 ps after optical excitation) serial femtosecond crystallography datasets.}
    \label{tab:datastats}
\end{table}

\newpage

\section{Average quantum-dot occupation}

\subsection{Absorption cross section and laser fluence}

We obtained the cross section $\sigma_{exc}$ at the excitation wavelength from published values of the intrinsic absorption coefficient $\mu_{i,335}$ of $\CPB$ quantum dots (QDs) at $335\:\mathrm{nm}$~\cite{Maes:2018}:
\begin{equation*}
\sigma_{exc} = \mu_{i,exc}\times V_{QD} = \left(\mu_{i,335}\frac{A_{exc}}{A_{335}}\right)\times V_{QD}
\end{equation*}
Here, $V_{QD}$ is the QD volume. For the given sample, we thus obtained:
\begin{equation*}
\sigma_{477}=6.16\:10^{-15}\:\mathrm{cm}^2 
\end{equation*}
We subsequently set the laser fluence $J$ such that the photon flux $\phi_{477}$ was equal to $1/\sigma_{477}$. Hence:
\begin{equation*}
    J = 67.6\:\mu\mathrm{J}\cdot\mathrm{cm}^{-2}
\end{equation*}

\subsection{Average quantum-dot occupation}

$\CPB$ quantum dots (QDs) feature a mixed confinement regime, with a 2-fold degenerate electron state and a localized hole state. For the QDs studied here, formation of one electron-hole pair reduces the absorption cross-section for resonant excitation to $\approx 40\%$ of the initial value.\cite{Geiregat:2024} Furthermore, the complete occupation of the electron states blocks additional absorption to form 3 electron-hole pairs. We therefore estimate the fraction of QDs having ($P_0$) no, ($P_1$) one and ($P_2$) two electron hole pairs using the following set of dynamic equations:
\begin{gather*}
    \frac{dP_0}{dt} = -\gamma_0 P_0 \\
    \frac{dP_1}{dt} = \gamma_0 P_0 - \gamma_1 P_1 \\
    \frac{dP_2}{dt} = \gamma_1 P_1 \\
\end{gather*}
We thus obtain:
\begin{gather*}
    P_0 = e^{-\gamma_0 t}\\
    P_1 = \frac{\gamma_0}{\gamma_0-\gamma_1}\left(e^{-\gamma_1 t} - e^{-\gamma_0 t}\right) \\
    P_2 = 1 - P_0 - P_1
\end{gather*}
For the fluence chosen, we have $\gamma_0 t = 1$ and $\gamma_1 t= 0.4$ at the end of each laser pulse. One thus obtains:
\begin{gather*}
    P_0 = 0.368 \\
    P_1 = 0.504 \\
    P_2 = 0.128 
\end{gather*}
Since the 3 ps delay between the optical pump and the X-ray probe is shorter than the lifetime of two electron-hole pairs in a single QD~\cite{Geiregat:2024}, we assume that the X-rays probe QDs with the above occupation probability. 
Assuming that two electron-hole pairs yield double the distortion as one electron-hole pair, we thus obtain that on average, $0.504 + 2\times 0.128 = 0.760$ electron/hole pairs are probed per pulse. 

\newpage

\section{Orthorhombic relaxation model}
\begin{figure}[h]
  \centering
  \begin{tabular*}{\linewidth}{@{\extracolsep{\fill}}ccc}
      \includegraphics[width=0.3\textwidth]{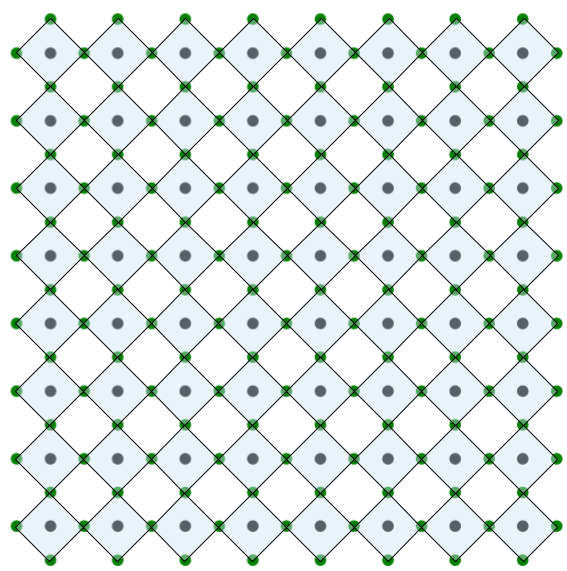} & \includegraphics[width=0.3\textwidth]{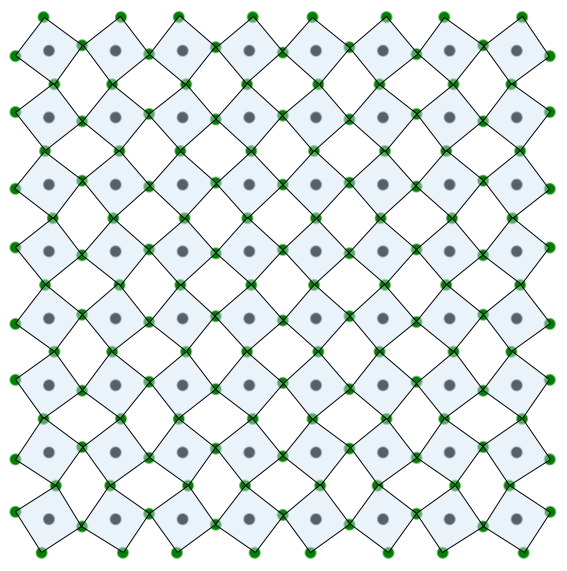} & \includegraphics[width=0.3\textwidth]{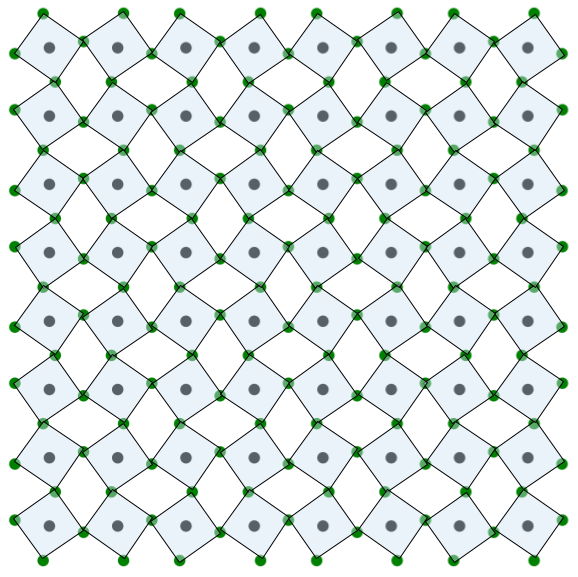} \\
      (a) & (b) & (c) \\
  \end{tabular*}
  \caption{Three different octahedral rotation models (a) No rotation, equivalent to the cubic structure. (b) Variable rotation, with smaller tilts near the centre of the particle. (c) Maximal rotation corresponding to the equilibrium orthorhombic structure.}
  \label{fig:orthotilts}
\end{figure}

The Jahn-Teller (J-T) distortion associated with the orthorhombic structure is associated with the rotation of the Br octahedra to break the electronic degeneracy. One possible polaronic lattice distortion model is to use this low-energy octahedral rotation mode. In this picture, the exciton causes the degree of octahedral tilt to vary as a function of distance from the centre of the polaron. In Fig.~\ref{fig:orthotilts}b, the orthorhombic J-T distortion is relaxed near the centre and approaches the equilibrium value further away. The two extreme cases are shown in Fig.~\ref{fig:orthotilts}a and c. 

The effect of such a variable tilt distortion is shown in Fig.~3c of the main text, where no peak shifts are observed, but only changes in the total intensities. Note that the effect is strongly exaggerated for the purposes of illustration in Fig.~\ref{fig:orthotilts}b.

\newpage

\section{Density functional theory analysis of lattice deformation in $\CPB$ quantum dots}

\subsection{Computational approach}

\subsubsection{General methodology}

To analyze the formation of a polaron in $\CPB$ QDs, we compared the relaxed QD geometry as obtained for the electronic ground state and the first excited state as predicted by density functional theory (DFT). These geometries were obtained by imposing either a total spin $S=0$, or a total spin $S=1$. Clearly, the latter will give the triplet state of the exciton, but the assumption is that the relaxation of the QD geometry will be highly similar for the singlet and triplet exciton. 
The DFT analysis provides the electron density for the valence electrons, and the $(x, y, z)$ coordinates of the different atoms. Using this information, 3D scattering patterns can be calculated for the ground state and the excited state by Fourier transformation, where the difference in patterns provides a result that can be compared directly with experimentally measured differences in diffraction intensity.

DFT calculations were implemented in CP2K, using the PBE functional. A fixed sequence was followed, in which first a hand-picked structure, see next section, was relaxed in the ground state. The resulting geometry was then used as an input for relaxing the excited state structure, which was defined by setting the spin multiplicity to 3 instead of 1. For both structures, the electron density was subsequently calculated using an energy calculation. The \texttt{.xyz} and \texttt{.cube} files providing the atomic coordinates and the electron density on a 3D grid were then taken as an input to determine the diffraction pattern for the ground and excited state. All spatial dimensions are expressed in units of the Bohr radius, $a_0 = 0.529$\AA.

\subsubsection{The $\CPB$ quantum dot models}

The analysis made use of different charge neutral $\CPB$ QD models with brute formula $\mathrm{Cs}_{200}\mathrm{Pb}_{125}\mathrm{Br}_{450}$. All models were cut as cubes from a bulk $\CPB$ crystal creating an inner framework of $5\times5\times5$ Pb atoms, and an outer framework of $6\times6\times6$ Cs atoms. To attain 200 Cs atoms, 16 vacancies were created in the outer Cs layer. For all models, the 8 Cs atoms from the cube corners were removed. For most other models, 8 additional Cs atoms were removed in different patterns from the cube edges. In one model, 4 Cs atoms were removed from edges, and 4 Cs atoms from facets. 

\subsection{Calculation of diffraction patterns}

\subsubsection{The diffraction pattern of the valence electrons}

The valence electron density was computed as a volumetric quantity on an equidistant 3D coordinate grid. To obtain the diffraction pattern with sufficient resolution in reciprocal space, the Fourier transform of the electron density was determined by calculating the Fourier integral across a limited region in reciprocal space around a given k-point. A reduction from a 3D to a 2D problem was achieved by first projecting the electron density on a given plane (in practice, the $xy$, $yz$ and $zx$ planes), after which the intersection of 3D reciprocal space with the corresponding 2D plane was obtained from a Fourier transform of the projected density:
\begin{equation}
F_{val}(q_1,q_2)=\int \rho_{12}(x_1, x_2)e^{i(q_1x_1+q_2x_2)}dx_1dx_2
\end{equation}
Here, $x_1$ and $x_2$ are the coordinates of the relevant coordinate plane, $\rho_{12}$ the projected electron density and $(q_1,q_2)$ the coordinates in the corresponding plane in reciprocal space.

\subsubsection{The diffraction pattern of the core electrons}

The contribution of the core electrons to the diffraction pattern was determined by considering each core atom as a delta-point scatterer, such that the diffraction amplitude $F_{core}(\bm{k})$ could be determined by a direct summation of the relevant phase factors:
\begin{equation}
F_{core}(\bm{q})=\sum_i N_i e^{i\bm{q}\cdot\bm{r}}
\end{equation}
Here, the index $i$ labels all different core atoms, and $N_i$ is the number of core electrons for each atom as used in the DFT calculation. By selecting $\bm{q}$ vectors using the same grid as for the valence electrons, a mutually compatible diffraction amplitude is obtained.  

\subsubsection{The diffraction intensity difference}

For a given 2D planar slice in reciprocal space, the 2D diffraction difference pattern $\Delta I_{diff}$ is determined from $F_{val}(q_1,q_2)$ and $F_{core}(q_1,q_2)$ according to:
\begin{equation}
\Delta I_{diff}(q_1,q_2) = \left|F_{ES,core}(q_1,q_2)+F_{ES,val}(q_1,q_2) \right|^2 - \left|F_{GS,core}(q_1,q_2)+F_{GS,val}(q_1,q_2) \right|^2 
\end{equation}

\subsection{Model nanocrystal relaxation}

\subsubsection{Energy changes upon relaxation}

\begin{figure}
    \centering
    \includegraphics[width=1\linewidth]{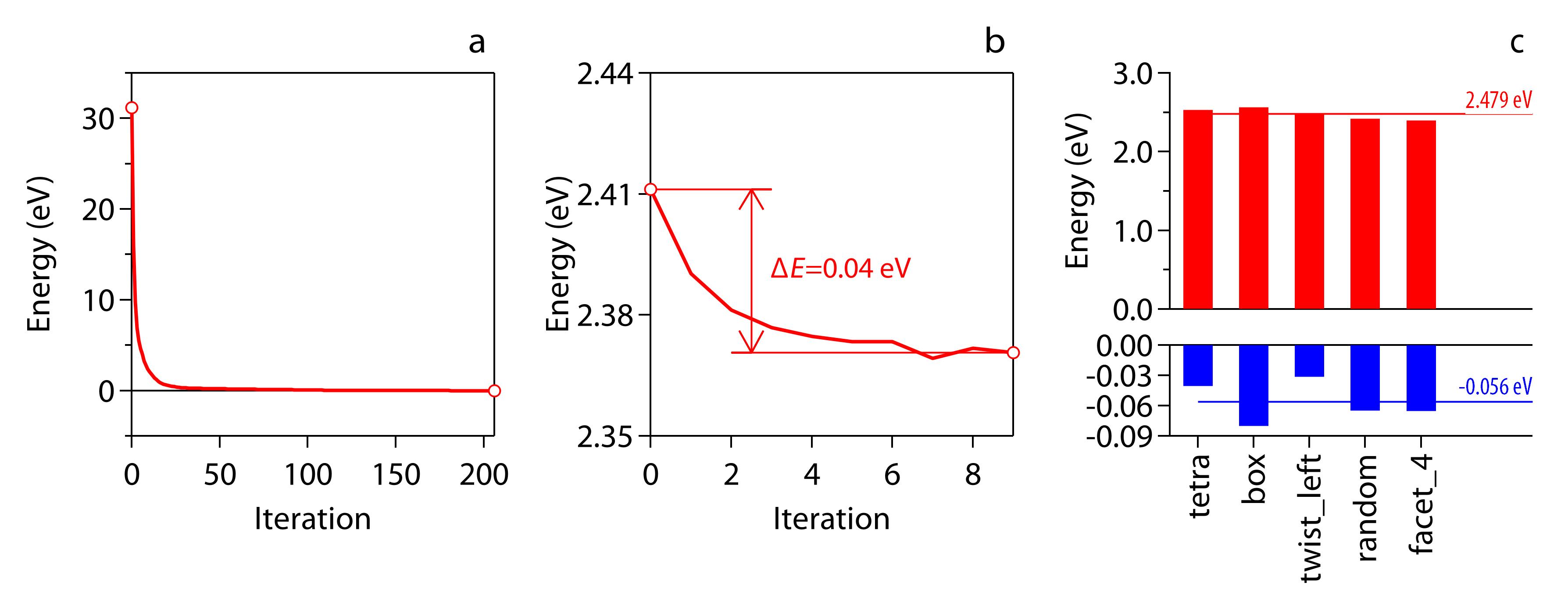}
    \caption{(a) Geometry optimization through relaxation of the ground state energy for the tetra NC, starting from the hand-picked structure. (b) Geometry optimization through relaxation of the excited-state energy for the tetra NC, starting from the relaxed ground-state structure. For (a) and (b), the energy of the relaxed ground state is taken as the energy reference. (c) Representation of (red) the energy difference between the relaxed excited state and the relaxed ground state and (blue) the relaxation energy for the excited state for the different model NCs. These quantities can be compared to the experimental photon energy of the band-edge emission and the Stokes shift.}
    \label{fig:SI_relaxation}
\end{figure}

Figure~\ref{fig:SI_relaxation}a represents the evolution of the total energy of the tetra NC during the geometry optimization of the ground state, starting from the hand-picked structure. Similar relaxation curves are obtained for all the model NCs analyzed. As shown in Figure~\ref{fig:SI_relaxation}b, starting from the ground-state structure, the geometry optimization for the excited state leads to a small but systematic energy relaxation. Averaged over the different model NCs, a relaxation energy of 0.056 eV is obtained, while the energy difference between the relaxed excited state and ground state amounts to 2.479 eV. In principle, the Stokes shift between the band-edge absorption and emission would amount to twice the relaxation energy, while the energy of the band-edge transition in absorption would correspond to the sum of the relaxed energy difference and the relaxation energy. 

\subsubsection{Ground state / excited state atom displacement field}

\begin{figure}
    \centering
    \includegraphics[width=1\linewidth]{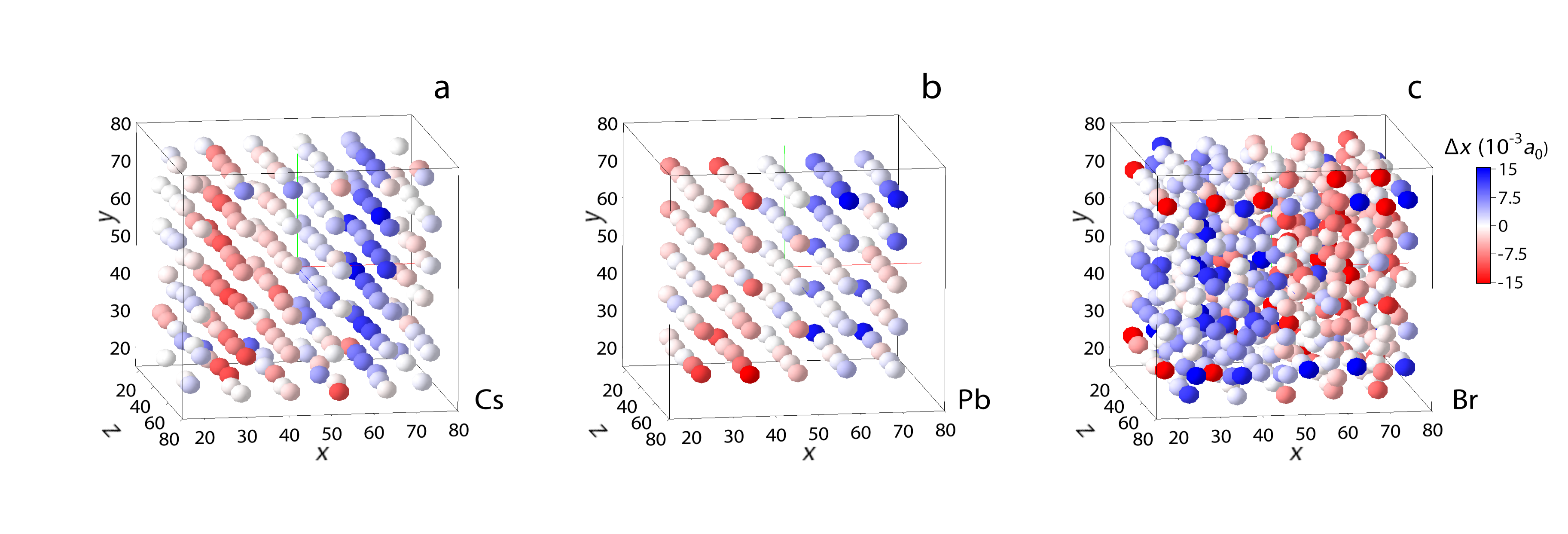}
    \caption{Color coded difference in $x$ position between the excited state and the ground state atomic positions for (a) Cs, (b) Pb and (c) Br, as obtained for the tetra NC. Blue means an atom has shifted in the direction of increasing $x$ in the excited state and red means an atom has shifted to decreasing $x$. The opposite shifts of Cs and Pb on the one hand, and Br on the other hand, are readily deduced from the images. The color bar is expressed in units of the Bohr radius $a_0$.}
    \label{fig:SI_shifts}
\end{figure}

As the core electrons will dominate the diffraction pattern, we first look at the changes in position of the core atoms when comparing the excited state and the ground state. Figure~\ref{fig:SI_shifts}a-c provides, as an example, the colour coded shifts along the $x$ direction of the Cs, Pb and Br atoms. Here, blue means a shift to the right (increasing $x$) and red means a shift to the left (decreasing $x$). As can be seen from the scatter plots, the Cs and Pb atoms at the left and the right of the NC feature shifts in line with their position along the $x$-axis, i.e., an outward displacement. The Br atoms, on the other hand, exhibit a shift opposite to their position along the $x$-axis, i.e., an inward displacement. A similar pattern emerges along the $y$ and $z$ directions, and appears for the four other NC structures analysed as well. 

\begin{figure}
    \centering
    \includegraphics[width=1\linewidth]{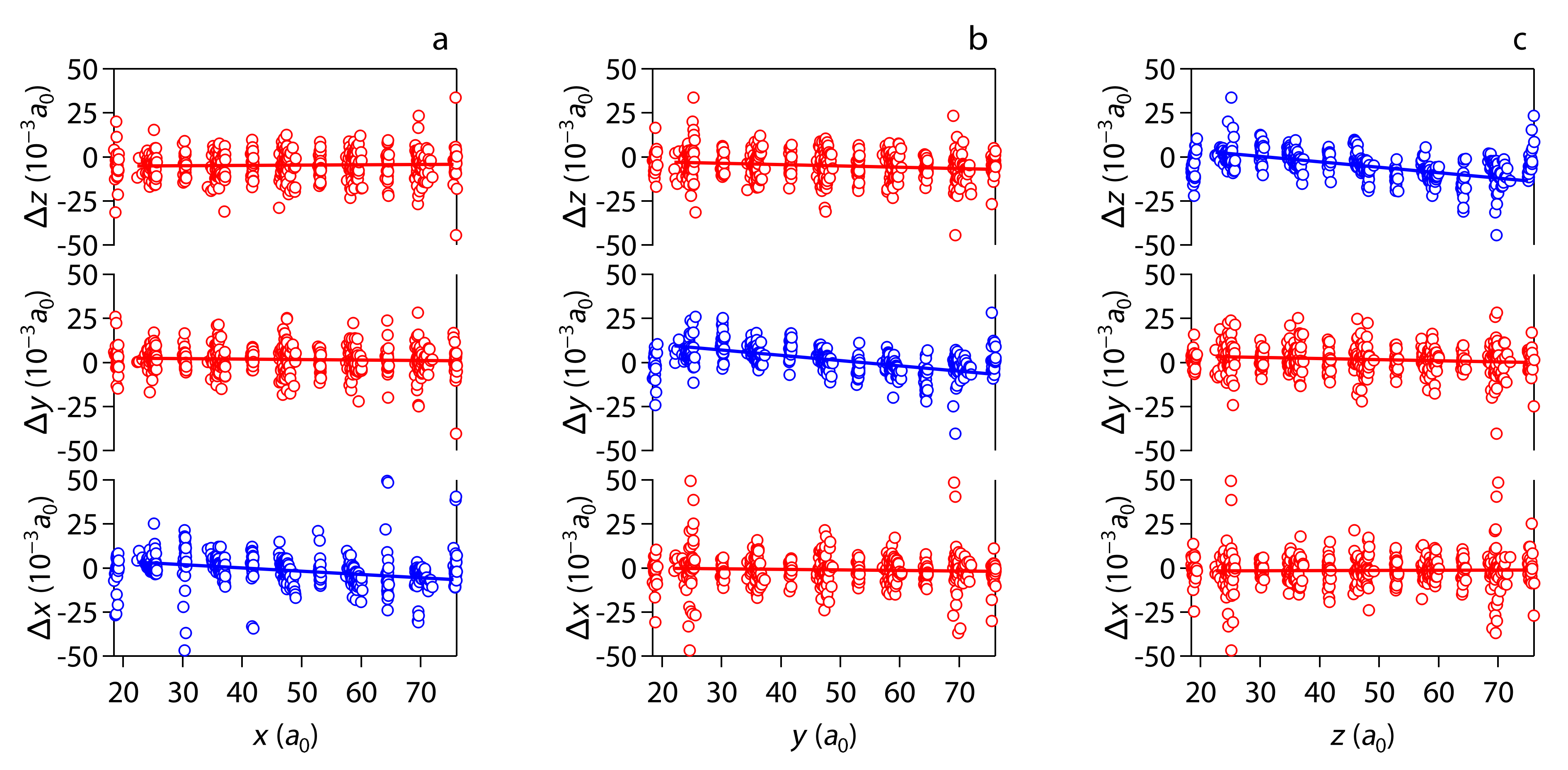}
    \caption{Shift of Br atoms in the $x$, $y$ and $z$ direction, plotted as a function of (a-c) the $x$, $y$ and $z$ coordinate of the Br atoms. All lines represent the result of a linear regression, excluding the Br atoms at the outer surfaces left and right.}
    \label{fig:SI_BrShifts}
\end{figure}

Figure~\ref{fig:SI_shifts} suggests that shifts along the x direction are mainly determined by the $x$ coordinate of a given atom, and vice versa for the other axis. Figure~\ref{fig:SI_BrShifts} provides the shifts of the Br position along $x$, $y$ and $z$, plotted as a function of the $x$, $y$ and $z$ coordinate, respectively. Invariably, a linear regression yields a significant correlation for the shift $\Delta x$ along $x$, shift $\Delta y$ along $y$, and $\Delta z$ along $z$, while all other combinations are either uncorrelated, or weakly correlated. Table~\ref{Tab:SI_shifts} provides for each regression line the slope, and the error on the slope. 

\begin{table}[tb]
\centering
\begin{tabular}{cccc}
\hline
Br Shift/coordinate & $x$ & $y$ & $z$ \\ \hline
$\Delta x$ & -190$\pm$28 & -25$\pm$25 & 18$\pm$24 \\ 
$\Delta y$ & -28$\pm$30  & -300$\pm$22 & -75$\pm$26 \\ 
$\Delta z$ & 9$\pm$31    & -66$\pm$27  & -311$\pm$21 \\ \hline
Cs shift/coordinate & x & y & z \\ \hline
$\Delta x$ & 325$\pm$25  & 8$\pm$34   & -12$\pm$34 \\
$\Delta y$ & -9$\pm$71   & 905$\pm$31 & 52$\pm$70 \\ 
$\Delta z$ & -16$\pm$74  & 73$\pm$71  & 920$\pm$31 \\ \hline
Pb shift/coordinate & $x$ & $y$ & $z$ \\ \hline
$\Delta x$ & 386$\pm$48  & 43$\pm$45  & -20$\pm$36 \\
$\Delta y$ & 74$\pm$74   & 302$\pm$50  & 106$\pm$35 \\ 
$\Delta z$ & 13$\pm$60   & 137$\pm$36  & 275$\pm$54 \\ \hline
\end{tabular}
\caption{Relative positional shift in a given direction (rows) for atoms along a given direction (columns). Shifts give the slope of linear fits to the atomic displacement, as shown in Fig.~\ref{fig:SI_BrShifts} for Br in parts per million of the interatomic distance.}
\label{Tab:SI_shifts}
\end{table}

\subsection{Nanocrystal diffraction patterns}

\subsubsection{Diffraction from the core electrons}

\begin{figure}
    \centering
    \includegraphics[width=\linewidth]{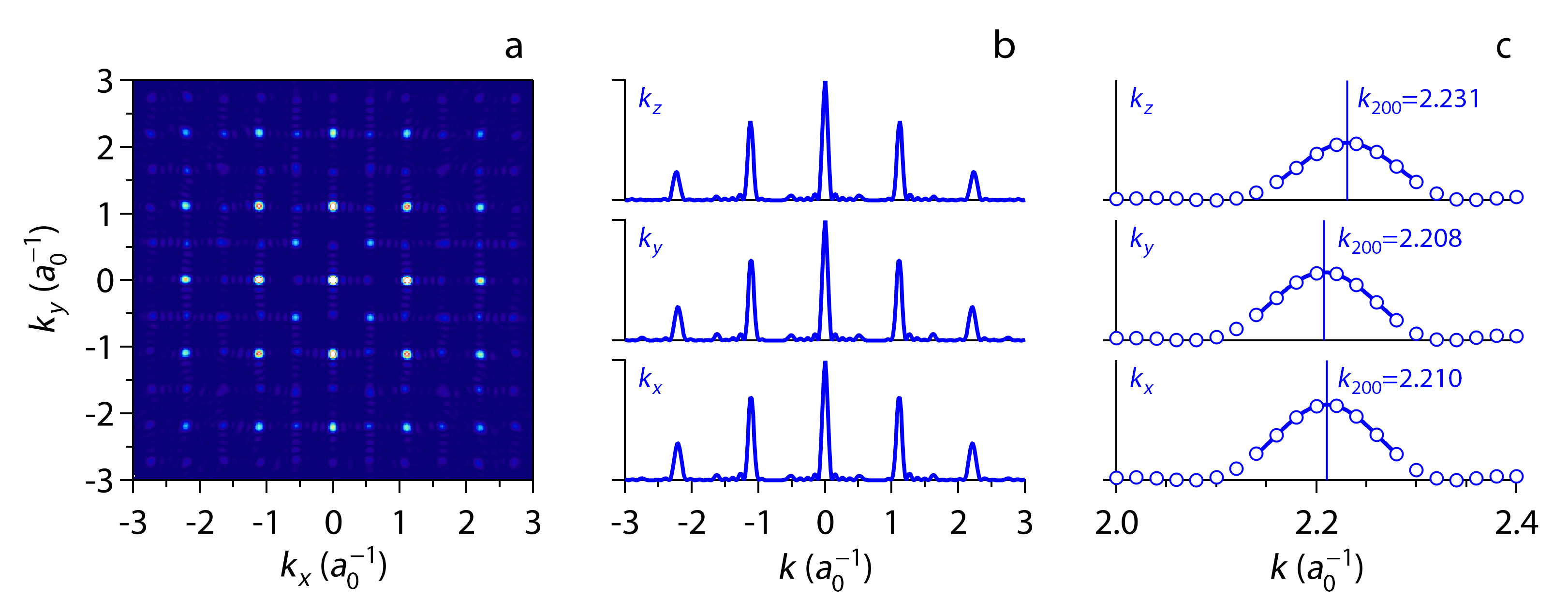}
    \caption{(a) kx-ky slice of the diffraction pattern of the core electrons in the tetra NC. (b) Line intensities along the different direction in reciprocal space as indicated. (c) Zoom on the line intensities around the (400) diffraction peak with a best fit of the central part to a Gaussian, for which the central wavenumber is indicated.}
    \label{fig:SI_CoreDiffraction}
\end{figure}

Figure~\ref{fig:SI_CoreDiffraction}a represents the diffraction pattern obtained for the relaxed ground state of the tetra NC. The pattern reflects the nearly cubic symmetry of the CsPbBr$_3$ QD lattice, featuring the most intense diffraction peaks for the (200), (220), and (400) directions. Note that the intensity at $k=0$ amounts to $31550^2$, i.e., the square of the total number of core electrons. A detailed view of the diffraction pattern along $k_x$, $k_y$, or $k_z$ shows that for this NC, the (400) peak attains a maximum at a slightly larger wavenumber along $k_z$ than along $k_x$ and $k_y$.

The intensity difference between the even 200 and 400 peaks and the odd 100 and 300 peaks results from the phase of the contributions of the different atoms to $F_\text{core}$. For even order peaks, all atoms within the unit cell—when perfectly positioned on the lattice points—contribute in phase. Hence, $F_\text{core}$ is the sum of the scattering from one Pb, one Cs, and three Br atoms per unit cell. For odd order peaks, there is a phase shift of $\pi$ between the scattering from the Cs and Br atoms at the edge of the unit cell, and the Pb and two Br atoms in the center of the unit cell considered along the (100) direction. Hence, $F_\text{core}$ is the difference of the contribution of one Pb and one Br, and one Cs atom per unit cell.

\subsection{Diffraction from the valence electrons}

\begin{figure}
    \centering
    \includegraphics{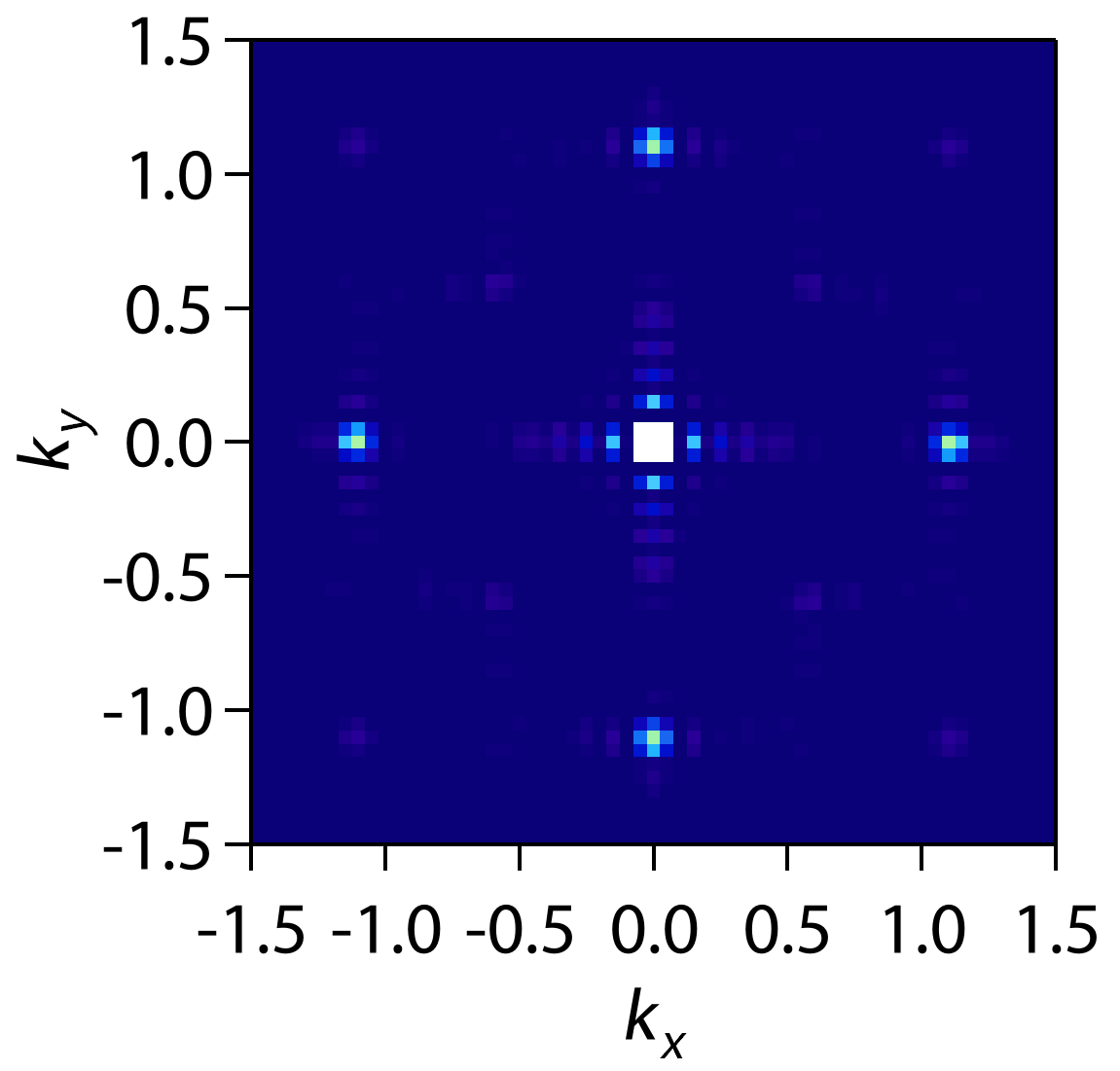}
    \caption{Low resolution $k_x-k_y$ slice of the diffraction pattern of the valence electrons in the tetra NC. }
    \label{fig:SI_ValenceDiffraction}
\end{figure}

Figure~\ref{fig:SI_ValenceDiffraction} represents a slice of the diffraction pattern obtained from the density of the valence electrons calculated at low resolution. Apart from $k=0$ diffraction, mostly the (200) diffraction peak is visible. The intensity at the center of the k=0 peak amounts to 29702500=$5450^2$, i.e., the square of the number of valence electrons. Overall, the contribution of the electron density distribution to the diffraction is far smaller than that of the core electrons. In line with the real space electron density, the (200) diffraction peak is dominant, but far less intense than the (200) diffraction from the core electrons.

\subsection{Excited State / Ground State Diffraction difference patterns}

\subsubsection{Detailed analysis of the (200) diffraction}

\begin{figure}[tb]
    \centering
    \includegraphics{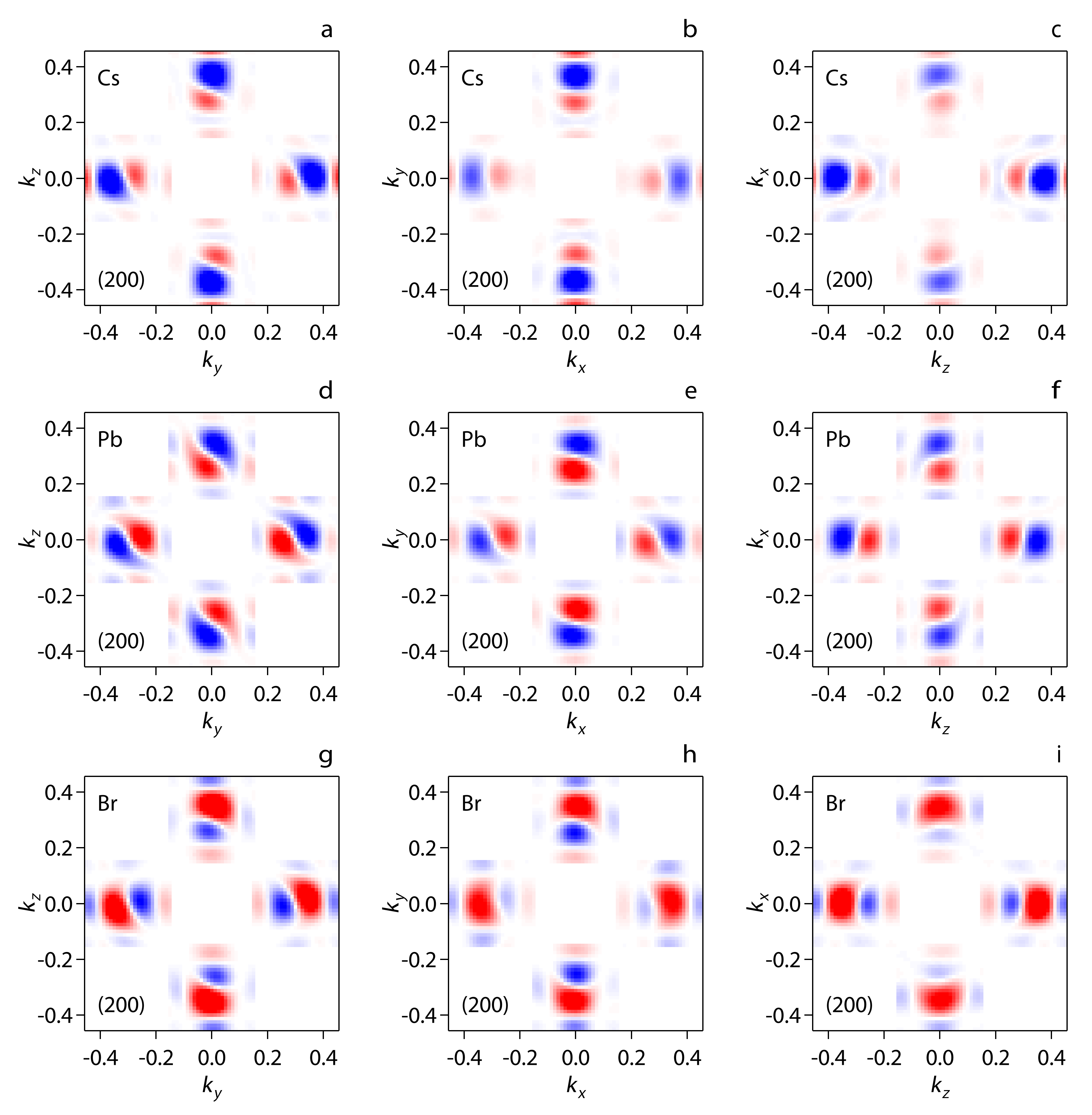}
    \caption{Single atom diffraction intensity difference simulated for (a) Cs-only, (b) Pb-only and (c) Br-only diffraction. Image plots from left to right represent different slices perpendicular to the 3 main axes in reciprocal space. Red colors represent an increased diffraction intensity in the excited state, blue colors a reduced diffraction intensity. Note that the (200) peaks have been shifted in reciprocal space from a center position at $k=1.1$ to $k=0.3$ to enhance the clarity of the patterns.}
    \label{fig:SI_SingleAtomDiffDiff}
\end{figure}

To develop an understanding as to how the atomic shifts lead to a difference between the diffraction pattern of the excited state and the ground state, Figure~\ref{fig:SI_SingleAtomDiffDiff} represents the simulated diffraction difference for the (200) peak, related to only Cs, only Pb or only the Br atoms. In line with the concomitant outward shift of Cs and Pb – which essentially increases the lattice parameter for the Cs and Pb sublattice – one sees that the diffraction related to Cs and Pb becomes more intense for smaller wavenumbers, and loses intensity for larger wavenumbers. The diffraction related to Br exhibits the opposite behavior, losing intensity at small wavenumbers and gaining intensity at large wavenumbers. Note that the shifts are not always entirely oriented towards the center of reciprocal space, an observation suggesting that some cross-correlation exists between, e.g, shifts $\Delta z$ along the y direction. For Br, such a secondary correlation can be deduced from Table~\ref{Tab:SI_shifts}.

\begin{figure}[tb]
    \centering
    \includegraphics{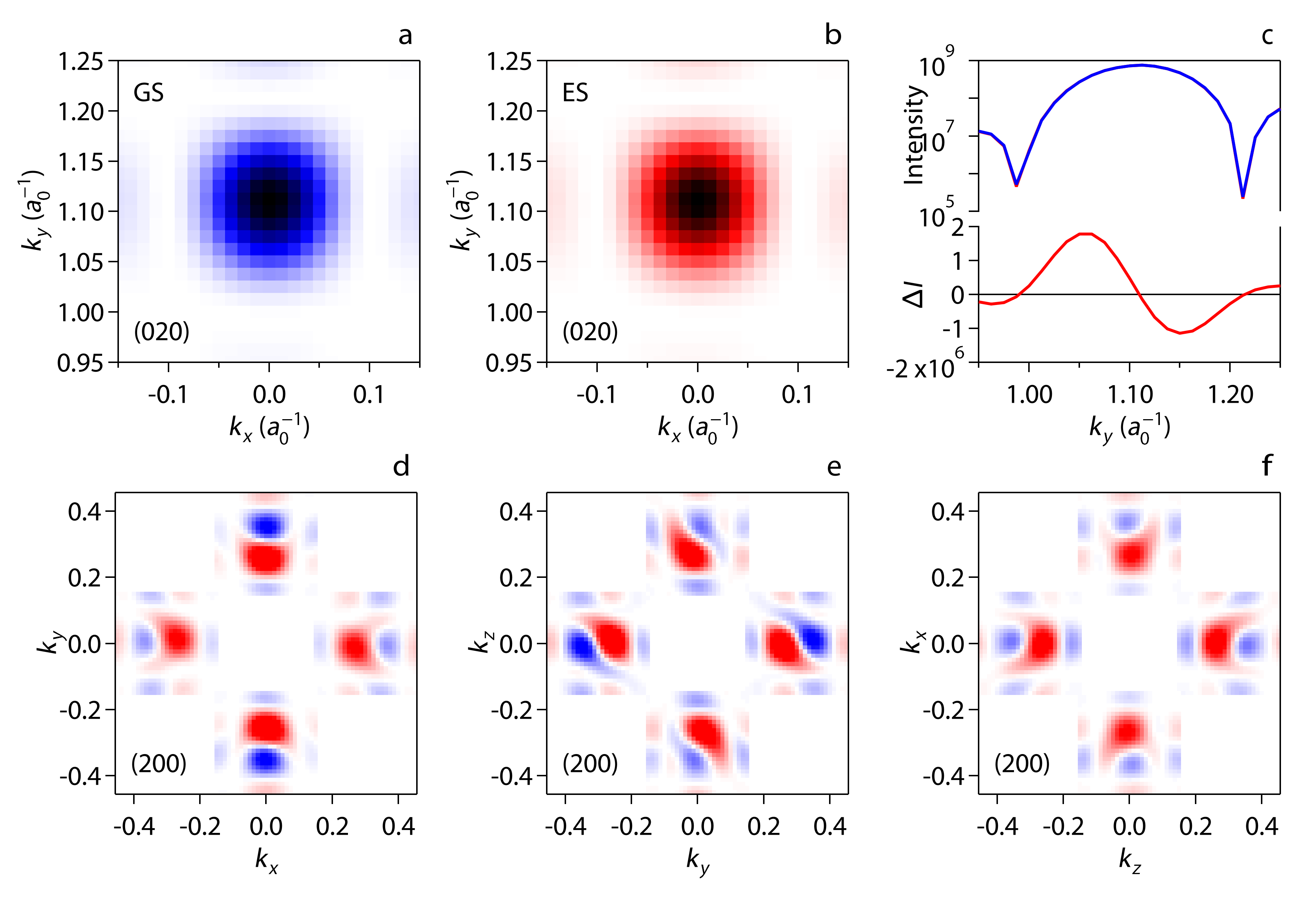}
    \caption{(a) Slice of the 020 diffraction peak through the $k_x-k_y$ plane for the relaxed ground-state of the tetra NC. (b) The same for the excited state. (c) (top) Diffraction intensity and (bottom) intensity difference of the (020) diffraction peak along the $k_y$ axis. The intensity profiles of the ground and excited states nearly overlap. (d-f) Diffraction intensity difference maps of the (200) peak for intersections with (d) the $k_x-k_y$ plane, (e) the $k_y-k_z$ plane and (f) the $k_z-k_x$ plane.}
    \label{fig:SI_DiffDiff200}
\end{figure}

Figure~\ref{fig:SI_DiffDiff200}a-b represent the total intensity of the (020) diffraction peak in the ground state and the excited state in the $k_x-k_y$ plane. As can be seen, both diffraction features are highly similar. Even so, the intensity difference shows a systematic variation with an increased intensity at smaller wavenumbers and a decreased intensity at larger wavenumbers after photo-excitation. Note that the intensity difference peaks at about 0.2\% of the peak intensity. Figure~\ref{fig:SI_DiffDiff200}d-f represents the intensity difference map of the (200) peak, counting all contributions from the core and the valence electrons. In line with the line section shown in Figure~\ref{fig:SI_DiffDiff200}c, one sees that, on the whole, the diffraction intensity increases at smaller wavenumbers, while the diffraction intensity decreases at larger wavenumbers. Referring to Figure~\ref{fig:SI_SingleAtomDiffDiff}, this implies that the changes to the diffraction pattern are dominated by the in-phase contributions from the shifts of the Cs and Pb atoms, and not the Br atoms. This result is not unexpected. For the (200) diffraction, the atoms at all lattice positions contribute to the diffraction patterns with the same phase. Hence, notwithstanding slight shifts relative to the lattice position, the diffraction intensity will be the sum of the intensity of the diffraction by the different atoms, such that a change in intensity will directly reflect changes in the atomic position. Since Cs and Pb represent 124 core electrons, while three Br atoms account for only 84, the overall intensity difference is dominated by the shift of the former atoms, not the latter. 

\subsubsection{Comparing different CsPbBr$_3$ model QDs}

\begin{figure}[h]
    \centering
    \includegraphics{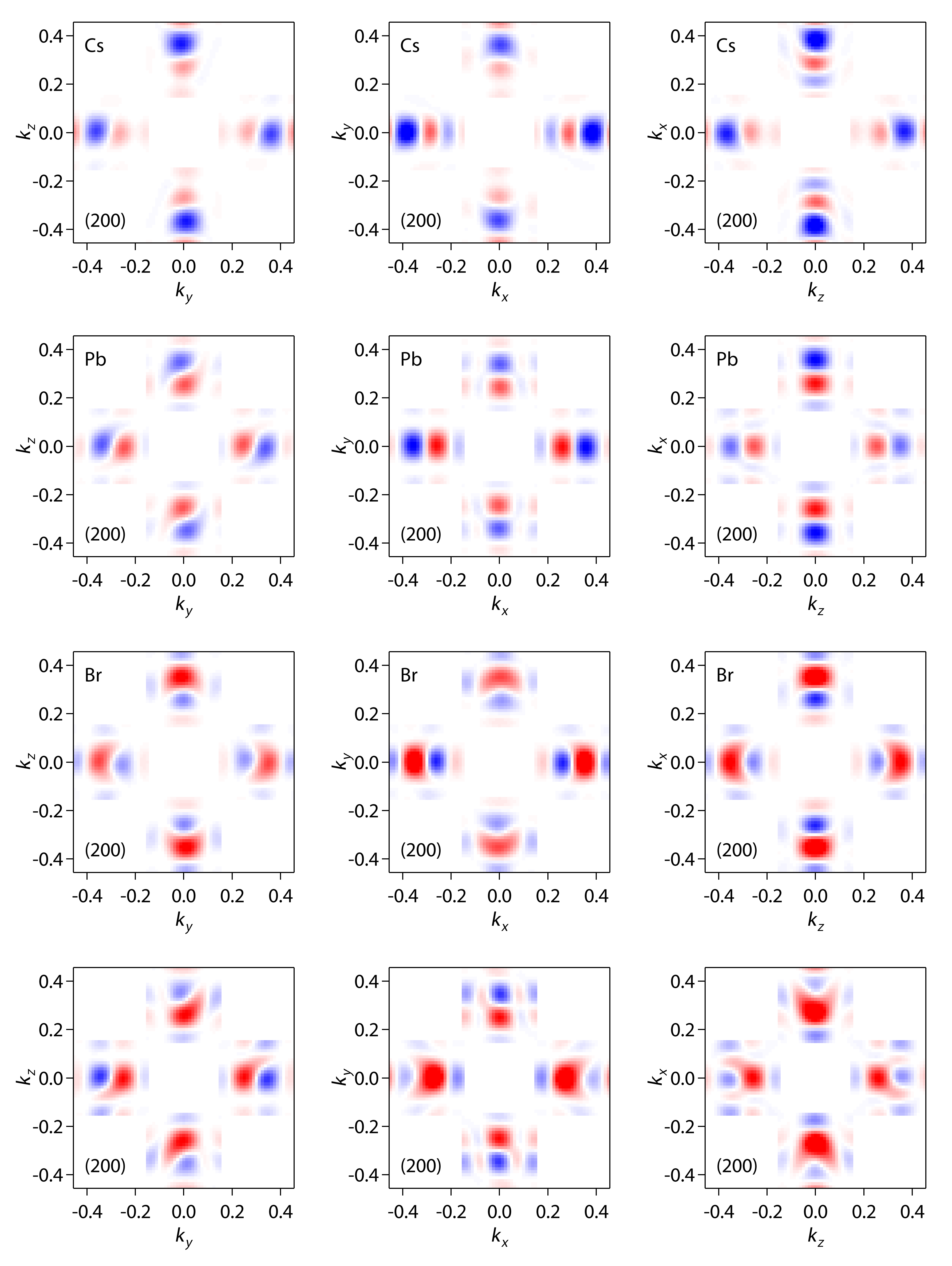}
    \caption{(3 top rows) Atom-selective diffraction intensity difference maps around the (200) peak and (bottom row) Total diffraction intensity difference around the (200) peak for the Facet4 QD.}
    \label{fig:SI_XRDFacet4}
\end{figure}

\begin{figure}[h]
    \centering
    \includegraphics{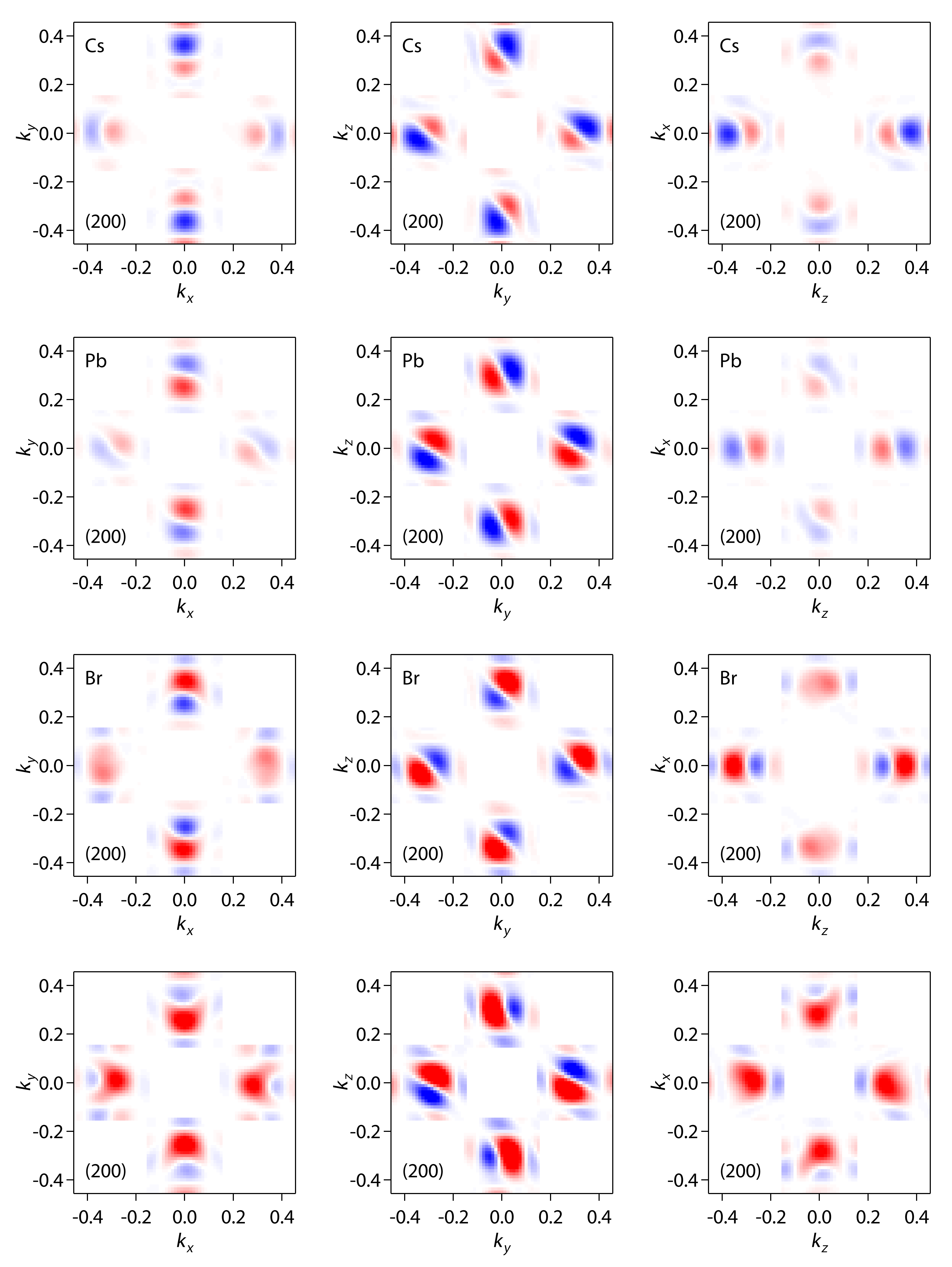}
    \caption{(3 top rows) Atom-selective diffraction intensity difference maps around the (200) peak and (bottom row) Total diffraction intensity difference around the (200) peak for the box NC.}
    \label{fig:SI_XRDBox}
\end{figure}

The symmetry of the model QD is affected by the arrangement of the Cs vacancies at the surface. In the case of the box NC, for example, the vacancies are organized such that the $x$-axis is reduced to a C2 symmetry axis. Opposite to the case of the tetra NC, this arrangement implies that the atom displacement cannot be described as a simple displacement towards or away from the center. Figures~\ref{fig:SI_XRDFacet4} and \ref{fig:SI_XRDBox} show the different slices through the 200 diffraction intensity differences, calculated for Cs, Pb and Br separately, and considering the total diffraction, for the case of the Facet4 and the Box NC. As can be seen, the Facet4 NC exhibits a diffraction intensity difference highly similar to the tetra NC, with features pointing towards the center of reciprocal space. The Box NC, on the other hand, features more complex diffraction intensity differences that reflect more involved atom displacement fields. On the other hand, the atom-selective patterns still highlight that the displacement of Br is opposite to the displacement of Cs and Pb. This point is a recurring feature of all the model NCs analyzed.

\newpage

\section{Restoring force tuning}

In the Random Hole Localization Model, the lattice displacements are assumed to be in the direction of the local electric field, with the negative Br ions moving opposite to the field. The magnitude of the equilibrium displacement, $\vec{d}$, for each of the atom types will depend on the strength of the restoring force, the effective charge on the atom and local electric field, such that
\begin{equation}
    k \vec{d} = q \vec{E}
\end{equation}
where $k$ is the \emph{spring constant} of the local harmonic potential.

Assuming a spherically symmetric harmonic potential in the near neighbourhood of the equilibrium position, these restoring strengths can be reduced to a single number for each atomic species.

For the special case of the centrally localized hole and a delocalized electron, we sampled these three parameters keeping the total squared magnitude constant, i.e. $(k_\mathrm{Cs}^2 + k_\mathrm{Pb}^2 + k_\mathrm{Br}^2)$. For each sampled condition, the differential diffraction was calculated and the peak shift magnitudes for the 200 and 110 peaks were compared. 

\begin{figure}
  \centering
  \begin{tabular}{cc}
       \includegraphics[width=0.45\textwidth]{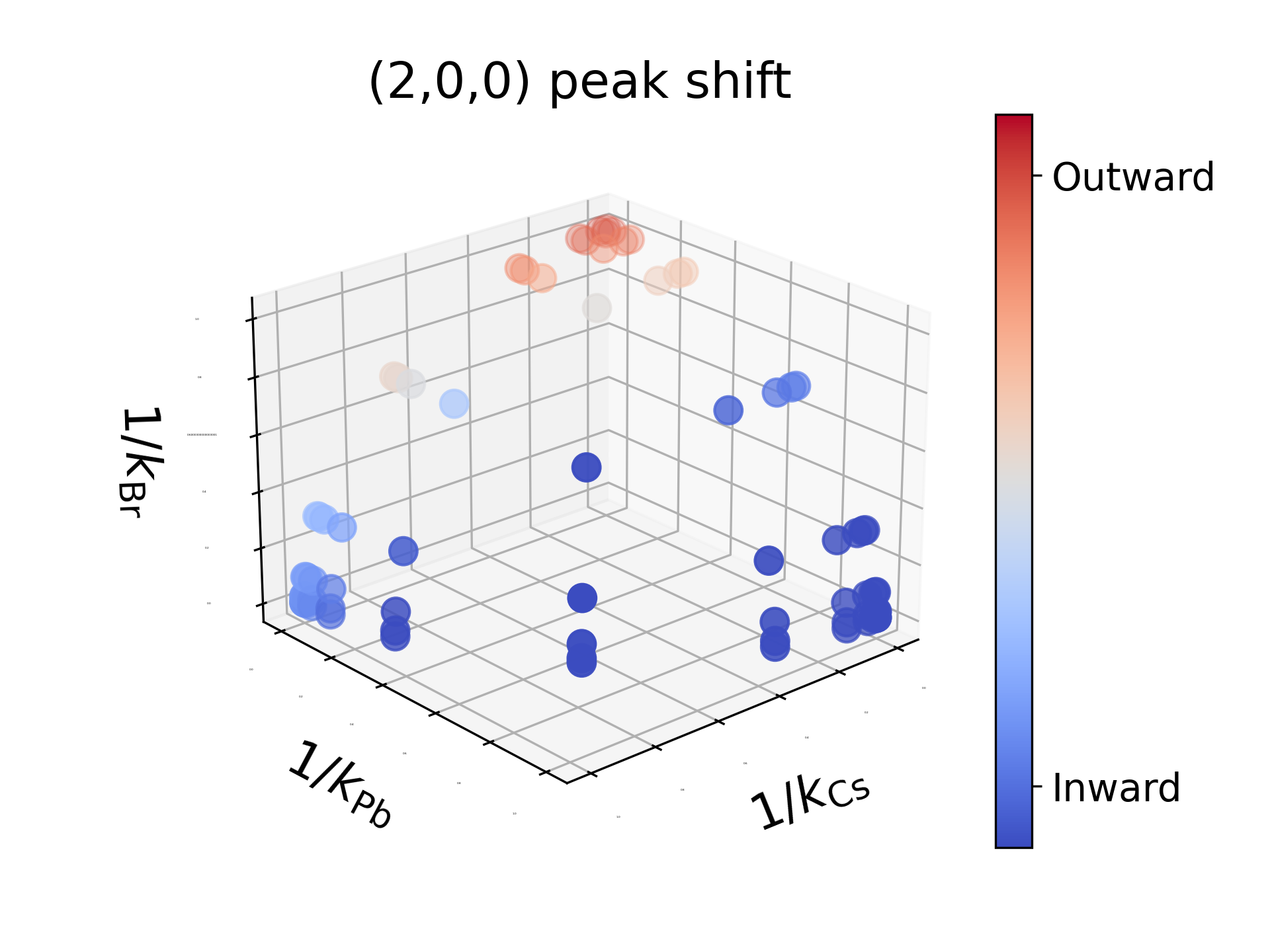} & \includegraphics[width=0.45\textwidth]{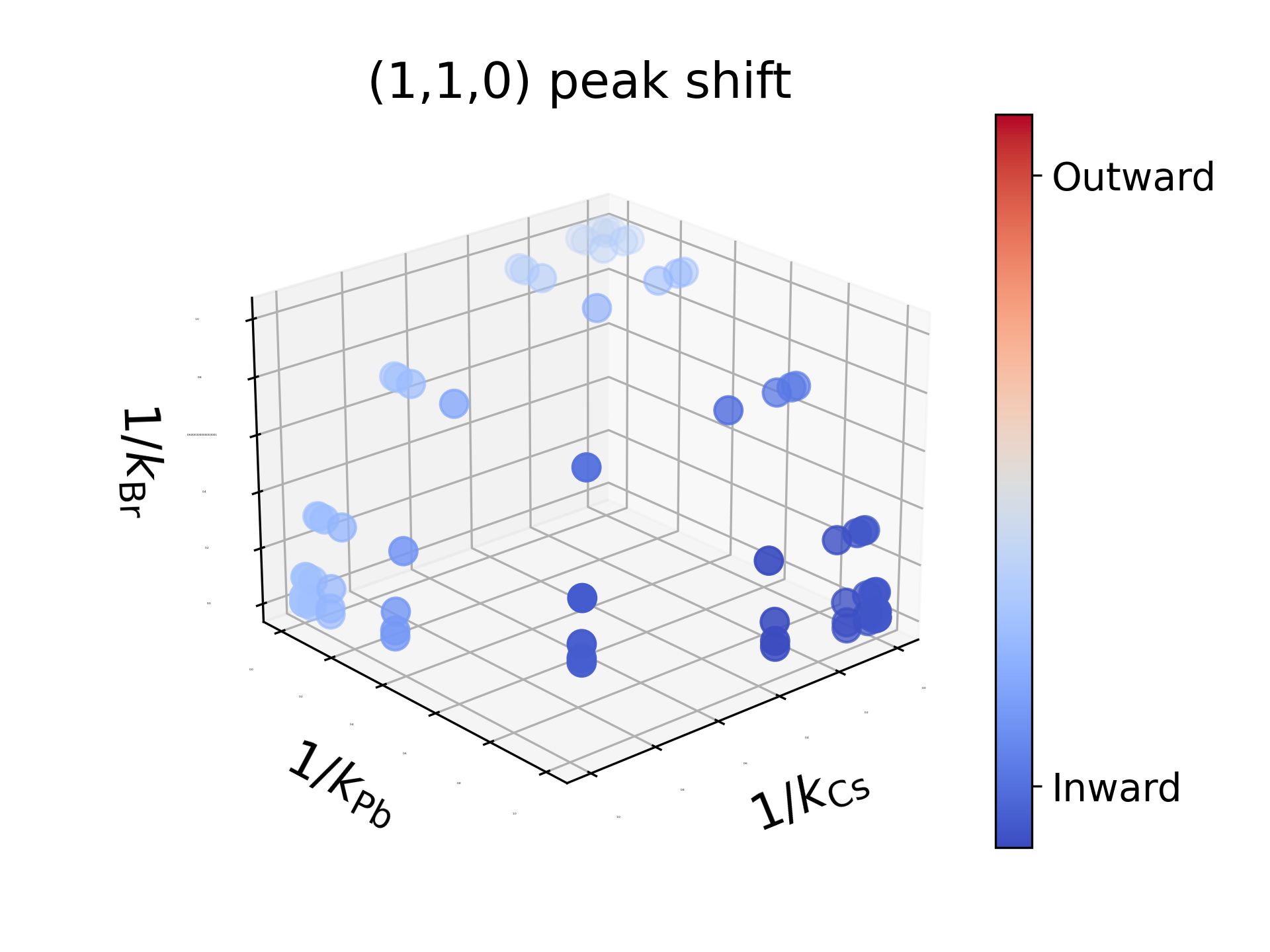} 
  \end{tabular}
  \caption{Peak shifts as a function of restoring force constants with a centrally localized hole charge. Each plot shows the peak shift magnitudes for different atomic restoring force strengths. Red and blue points indicate outward and inward shifts, respectively. In the experimental data, the 200 peak shifts inward and the 110 peak shifts outward upon optical pumping.}
  \label{fig:kscan}
\end{figure}

As one can see in Fig.~\ref{fig:kscan}, no choice of relative restoring strength, $k$, values generate an outward peak shift for the (110) peak. Thus, a centrally localized hole charge cannot account for the observed (110) shift. 

For the (200) peak, a relatively weak Br restoring force is required to produce an outward shift. However, this effect reverses for outwardly biased hole positions, wherein a weak Br $k$-value produces an inward shift. This is also true for the (400) peak and reproduces the final result shown in Fig.~5c of the main text.

\bibliography{refs}